\documentclass[final,twocolumn]{IEEEtran}

\usepackage[cmex10]{amsmath}
\usepackage{latexsym}
\usepackage{graphics}
\usepackage{amssymb}
\usepackage{amsthm}

\usepackage{cite}
\usepackage{array}
\usepackage{graphicx}
\usepackage[linesnumbered,ruled,norelsize,vlined]{algorithm2e}

\usepackage{url}
\usepackage{color}
\usepackage{subfigure}

\def\vectorize{\mathrm{vec}}

\def\kron{\otimes}
\def\tr{\mathrm{tr}}

\def\diag{\mathrm{diag}}
\def\vectorize{\mathrm{vec}}
\newcommand{\MMSE}[1]{\widehat{#1}_{\mathrm{MMSE}}}
\newcommand{\PEACH}[1]{\widehat{#1}_{\mathrm{PEACH}}}
\newcommand{\WPEACH}[1]{\widehat{#1}_{\mathrm{W}\textrm{-}\mathrm{PEACH}}}
\newcommand{\MVU}[1]{\widehat{#1}_{\mathrm{MVU}}}

\newcommand{\vect}[1]{\mathbf{#1}}

\newcommand{\minimize}[1]{{\underset{{#1}}{\mathrm{minimize}}}}

\theoremstyle{plain}
\newtheorem{remark}{Remark}
\newtheorem{theorem}{Theorem}

\newtheorem{proposition}{Proposition}
\newtheorem{lemma}{Lemma}

\IEEEoverridecommandlockouts

\begin{document}

\title{Low-Complexity Polynomial Channel Estimation in\\ Large-Scale MIMO with Arbitrary Statistics}

\author{Nafiseh Shariati,~\IEEEmembership{Student Member,~IEEE,}
        Emil~Bj\"ornson,~\IEEEmembership{Member,~IEEE,}
        Mats~Bengtsson,~\IEEEmembership{Senior~Member,~IEEE,}
        and~M\'erouane~Debbah,~\IEEEmembership{Senior~Member,~IEEE}
\thanks{Copyright (c) 2013 IEEE. Personal use of this material is permitted. However, permission to use this material for any other purposes must be obtained from the IEEE by sending a request to pubs-permissions@ieee.org.}
\thanks{N.~Shariati and M.~Bengtsson are with the Department of Signal Processing, ACCESS Linnaeus Centre, KTH Royal Institute of Technology, Stockholm, Sweden (e-mail: \{nafiseh, mats.bengtsson\}@ee.kth.se).}%
\thanks{E. Bj{\"o}rnson was with the Alcatel-Lucent Chair on Flexible Radio, Sup{\'e}lec, Gif-sur-Yvette, France, and with the Department of Signal Processing, KTH Royal Institute of Technology, Stockholm, Sweden. He is currently with the Department of Electrical Engineering (ISY), Link{\"o}ping University, Link{\"o}ping, Sweden (email: emil.bjornson@liu.se).}
\thanks{M.~Debbah is with the Alcatel-Lucent Chair on Flexible Radio, SUPELEC, Gif-sur-Yvette, France (e-mail:merouane.debbah@supelec.fr).}%
\thanks{This work was presented in part at IEEE Symposium on Personal, Indoor, Mobile and Radio Communications (PIMRC), London, UK, Sept.~2013. \cite{Shariati2013a}}%
\thanks{E.~Bj\"ornson is funded by the International Postdoc Grant 2012-228 from The Swedish Research Council. This research has been supported by the ERC Starting Grant 305123 MORE (Advanced Mathematical Tools for Complex Network Engineering).}
}

\maketitle

\begin{abstract}
This paper considers pilot-based channel estimation in large-scale multiple-input multiple-output (MIMO) communication systems, also known as ``massive MIMO'', where there are hundreds of antennas at one side of the link. Motivated by the fact that computational complexity is one of the main challenges in such systems, a set of low-complexity Bayesian channel estimators, coined \emph{Polynomial ExpAnsion CHannel (PEACH) estimators}, are introduced for arbitrary channel and interference statistics. While the conventional minimum mean square error (MMSE) estimator has cubic complexity in the dimension of the covariance matrices, due to an inversion operation, our proposed estimators significantly reduce this to  square complexity by approximating the inverse by a $L$-degree matrix polynomial. The coefficients of the polynomial are optimized to minimize the mean square error (MSE) of the estimate.

We show numerically that near-optimal MSEs are achieved with low polynomial degrees. We also derive the exact computational complexity of the proposed estimators, in terms of the floating-point operations (FLOPs), by which we prove that the proposed estimators outperform the conventional estimators in large-scale MIMO systems of practical dimensions while providing a reasonable MSEs. Moreover, we show that $L$ needs not scale with the system dimensions to maintain a certain normalized MSE. By analyzing different interference scenarios, we observe that the relative MSE loss of using the low-complexity PEACH estimators is smaller in realistic scenarios with pilot contamination. On the other hand, PEACH estimators are not well suited for noise-limited scenarios with high pilot power; therefore, we also introduce the low-complexity \emph{diagonalized estimator} that performs well in this regime. Finally, we also investigate numerically how the estimation performance is affected by having imperfect statistical knowledge. High robustness is achieved for large-dimensional matrices by using a new covariance estimate which is an affine function of the sample covariance matrix and a regularization term.
\end{abstract}

\begin{IEEEkeywords}
Channel estimation, large-scale MIMO, polynomial expansion, pilot contamination, spatial correlation.
\end{IEEEkeywords}

\IEEEpeerreviewmaketitle

\section{Introduction}

MIMO techniques can bring huge improvements in spectral efficiency to wireless systems, by increasing the spatial reuse through spatial multiplexing \cite{Lozano2010a}. While $8 \times 8$ MIMO transmissions have found its way into recent communication standards, such as LTE-Advanced \cite{Holma2012a}, there is an increasing interest from academy and industry to equip base stations (BSs) with much larger arrays with several hundreds of antenna elements \cite{Marzetta2010a,Jose2011b,Hoydis2013a,Rusek2013a,Larsson2014a,Baldemair2013a}. Such large-scale MIMO, or ``massive MIMO'', techniques can give unprecedented spatial resolution and array gain, thus enabling a very dense spatial reuse that potentially can keep up with the rapidly increasing demand for wireless connectivity and need for high energy efficiency.

The antenna elements in large-scale MIMO can be either collocated in one- or multi-dimensional arrays or distributed over a larger area (e.g., on the facade or the windows of buildings) \cite{Larsson2014a}. Apart from increasing the spectral efficiency of conventional wireless systems, which operate at carrier frequencies of one or a few GHz, the use of massive antenna configurations is also a key enabler for high-rate transmissions in mm-Wave bands, where there are plenty of unused spectrum today \cite{Baldemair2013a}. In particular, the array gain of large-scale MIMO mitigates the large propagation losses at such high frequencies and 256 antenna elements with half-wavelength minimal spacing can be packed into $6 \times 6 \,\, \mathrm{cm}$ at 80 GHz \cite{Baldemair2013a}.

The majority of previous works on large-scale MIMO (see \cite{Marzetta2010a,Jose2011b,Hoydis2013a,Rusek2013a,Larsson2014a} and references therein) considers scenarios where BSs equipped with many antennas communicate with single-antenna user terminals (UTs). While this assumption allows for closed-form characterizations of the asymptotic throughput (when the number of antennas and UTs grow large), we can expect practical UTs to be equipped with multiple antennas as well---this is indeed the case already in LTE-Advanced \cite{Holma2012a}. However, the limited form factor of terminals typically allows for fewer antennas than at the BSs, but the number might still be unconventionally large in mm-Wave communications.

A major limiting factor in large-scale MIMO is the availability of accurate instantaneous channel state information (CSI). This is since high spatial resolution can only be exploited if the propagation environment is precisely known. CSI is typically acquired by transmitting predefined pilot signals and estimating the channel coefficients from the received signals \cite{Yin2013a,Mueller2013a,Kay1993a,Kotecha2004a,Liu2007a,Bjornson2010a}. The pilot overhead is proportional to the number of transmit antennas, thus it is commonly assumed that the pilots are sent from the array with the smallest number of antennas and used for transmission in both directions by exploiting channel reciprocity in time-division duplex (TDD) mode.

The instantaneous channel matrix is acquired from the received pilot signal by applying an appropriate estimation scheme. The Bayesian MMSE estimator is optimal if the channel statistics are known \cite{Kay1993a,Kotecha2004a,Liu2007a,Bjornson2010a,Shariati2014a}, while the minimum-variance unbiased (MVU) estimator is applied otherwise \cite{Kay1993a}. These channel estimators basically solve a linear system of equations, or equivalently multiply the received pilot signal with an inverse of the covariance matrices. This is a mathematical operation with cubic computational complexity in the matrix dimension, which is the product of the number of antennas at the receiver (at the order of 100) and the length of the pilot sequence (at the order of 10). Evidently, this operation is extremely computationally expensive in large-scale MIMO systems, thus the MMSE and MVU channel estimates cannot be computed within a reasonable period of time. The high computational complexity can be avoided under propagation conditions where all covariance matrices are diagonal, but large-scale MIMO channels typically have a distinct spatial channel correlation due to insufficient antenna spacing and richness of the propagation environment \cite{Rusek2013a}. The spatial correlation decreases the estimation errors \cite{Bjornson2010a}, but only if an appropriate estimator is applied.
Moreover, the necessary pilot reuse in cellular networks creates spatially correlated inter-cell interference, known as \emph{pilot contamination}, which reduces the estimation performance and spectral efficiency \cite{Jose2011b,Hoydis2013a,Rusek2013a,Yin2013a,Mueller2013a}.

\emph{Polynomial expansion (PE)} is a well-known technique to reduce the complexity of large-dimensional matrix inversions \cite{Moshavi1996a}. Similar to classic Taylor series expansions for scalar functions, PE approximates a matrix function by an $L$-degree matrix polynomial. PE has a long history in the field of signal processing for multiuser detection/equalization, where both the decorrelating detector and the linear MMSE detector involve matrix inversions \cite{Moshavi1996a,Lei1998a,Muller2001a,Honig2001a,Sessler2005a,Hoydis2011d}. PE-based detectors are versatile since the structure enables simple multistage/pipelined hardware implementation \cite{Moshavi1996a} using only additions and multiplications. The degree $L$ basically describes the accuracy to which the inversion of each eigenvalue is approximated, thus the degree needs not scale with the system dimensions to achieve near optimal performance \cite{Honig2001a}. Instead, $L$ is simply selected to balance between computational complexity and detection performance. A main problem is to select the coefficients of the polynomial to achieve high performance at small $L$; the optimal coefficients are expensive to compute \cite{Moshavi1996a}, but alternatives based on appropriate scalings \cite{Lei1998a,Sessler2005a,Josse2008a} and asymptotic analysis \cite{Muller2001a,Hoydis2011d} exist. Recently, PE has also been used to reduce the precoding complexity in large-scale MIMO systems \cite{Zarei2013a,Mueller2014a,Kammoun2014b}, and high performance was achieved by optimizing the matrix polynomials using asymptotic analysis.

The optimization of the polynomial coefficients is the key to high performance when using PE. Since the system models and performance metrics are fundamentally different in multiuser detection and precoding, the derivation of optimal and low-complexity suboptimal coefficients become two very different problems in these two applications. In this paper, we consider a new signal processing application for PE, namely pilot-based estimation of MIMO channels. We apply the PE technique to approximate the MMSE estimator and thereby obtain a new set of low-complexity channel estimators that we coin \emph{Polynomial ExpAnsion CHannel (PEACH)} estimators.\footnote{After the submission of this paper, we became aware of the concurrent work of \cite{Chen2013a} which also applies PE to reduce the complexity of MMSE estimation. However, orthogonal frequency division multiplexing (OFDM) systems with a large number of subcarriers are considered in \cite{Chen2013a}, while large-scale single-carrier MIMO systems are our focus. This makes the system models, analysis, and results non-overlapping.} A main contribution of the paper is to optimize the coefficients of the polynomial to yield low MSE at any fixed polynomial degree $L$, while keeping the low complexity. The PEACH estimators are evaluated under different propagation/interference conditions and show remarkably good performance at low polynomial degrees.  An important property is that $L$ needs not scale with the number of antennas to maintain a fixed normalized MSE loss (as compared to MMSE estimation). However, $L$ should increase with the transmit power to keep a fixed loss, while it can actually be decreased as the interference becomes stronger. The computational complexity of the PEACH estimators and conventional MMSE/MVU estimators are compared analytically. This reveals that the proposed estimators have smaller complexity exponents. The numerical results confirm that much fewer FLOPs are required to compute the PEACH estimators in large-scale MIMO systems of practical dimensions. Finally, the \emph{diagonalized estimator} is introduced with even lower complexity and it is shown in which scenarios it is suitable.

\subsection{Outline}

The organization of this paper is as follows. In Section \ref{section:Problem_Formulation}, we describe the system model and formulate the problem of estimating channel coefficients for a large-scale MIMO communication system where the computational complexity is a major issue. Following the Bayesian philosophy, we propose a set of low-complexity estimators in Section \ref{section:Low_complexity estimators} and provide an exact complexity analysis. In Section  \ref{section:Numerical_Evaluation}, we numerically evaluate the performance of the proposed estimators in different interference scenarios where comparison is performed with respect to conventional estimators. Finally, conclusions are drawn in Section \ref{section:Conclusions}.

\subsection{Notation}

Boldface (lower case) is used for column vectors, $\vect{x}$, and
(upper case) for matrices, $\vect{X}$. Let  $\vect{X}^T$,
$\vect{X}^H$, and $\vect{X}^{-1}$ denote the transpose, the conjugate
transpose, and the inverse of $\vect{X}$, respectively. The Kronecker product of $\vect{X}$ and $\vect{Y}$ is
denoted $\vect{X} \otimes \vect{Y}$, $\vectorize(\vect{X})$ is the
 vector obtained by stacking the columns of $\vect{X}$,
$\tr( \vect{X} )$ denotes the trace, $\|\vect{X}\|_F$ is the Frobenius norm, and $\|\vect{X}\|_2$ is the spectral norm. The notation $\triangleq$ denotes definitions, while the big-$\mathcal{O}$ notation $\mathcal{O}(M^x)$ describes that the complexity is bounded by $ C M^x$ for some $0 < C < \infty$. A circularly symmetric complex Gaussian random vector $\vect{x}$ is denoted $\vect{x} \sim \mathcal{CN}(\bar{\vect{x}},\vect{Q})$, where $\bar{\vect{x}}$ is the mean and $\vect{Q}$ is the covariance matrix.

\begin{figure}
   \begin{center}
   \includegraphics[width=.9\columnwidth]{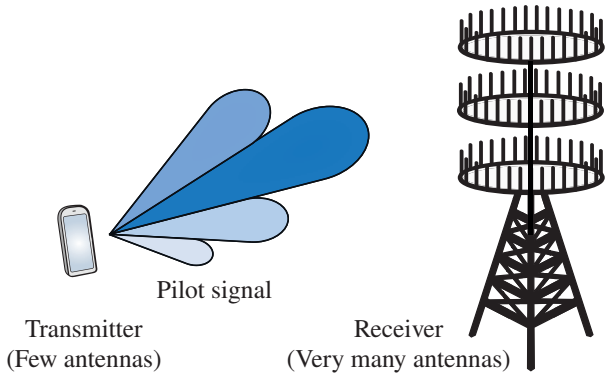}\\
 \caption{Illustration of pilot signaling in a large-scale $N_t \times N_r$ MIMO system, where typically $N_r \gg N_t$. The complexity of conventional channel estimators is very large in these systems, which calls for low-complexity alternatives.}
   \label{fig:problem_illustration}
   \end{center}
\end{figure}

\section{Problem Formulation}
\label{section:Problem_Formulation}

We consider a MIMO channel where the receiver and the transmitter are equipped with $N_r$ and $N_t$ number of antennas, respectively.
This can be one of the links in a multi-cell multi-user network of arbitrary size. The problem of estimating the instantaneous MIMO channel coefficients for a quasi-static flat-fading channel $\vect{H} \in \mathbb{C}^{N_r \times N_t}$ is investigated. The channel matrix $\mathbf{H}$ is modeled as Rician fading with $\vectorize(\vect{H}) \sim \mathcal{CN}(\vectorize(\vect{\bar{H}}),\vect{R})$ where the non-zero mean matrix $\vect{\bar{H}}$ implies that there might be line-of-sight propagation and the channel covariance matrix $\vect{R} \in \mathbb{C}^{N_t N_r \times N_t N_r}$ is positive semi-definite. Observe that $\vect{R}$ is generally \emph{not} a scaled identity matrix, but describes the spatial propagation environment. In order to estimate the channel coefficients, we exploit \textit{pilot} signals similar to \cite{Kotecha2004a,Liu2007a,Bjornson2010a}. This means that the transmitter sends the columns of a fixed predefined pilot matrix $\vect{P} \in \mathbb{C}^{N_t \times B}$ over $B$ channel uses; see Fig.~\ref{fig:problem_illustration}. The integer $B$ is the length of the pilot sequence and usually satisfies $B \geq N_t$.\footnote{Pilot sequences shorter than $N_t$ are optimal in highly correlated channels where the pilot matrix $\vect{P}$ is tailored to the channel and interference statistics \cite{Bjornson2010a}. The analysis herein permits any $B \geq 1$,  but we stress that $B \geq N_t$ is the case of main interest. This is due to the fact that pilot matrix optimization is cumbersome in large-scale MIMO systems since the transmitter and receiver need to acquire the same statistical information to agree on the pilot matrix.}

During the pilot signaling, the received matrix $\vect{Y} \triangleq [\vect{y}(1), \cdots, \vect{y}(B)]$ equals
\begin{equation} \label{eq_pilotmodel}
\vect{Y} = \vect{H} \vect{P} + \vect{N}
\end{equation}
where the disturbance $\vect{N} \in \mathbb{C}^{N_r \times B}$ is assumed to be circularly-symmetric complex Gaussian distributed and modeled as $\vectorize(\vect{N}) \sim \mathcal{CN}(\vectorize(\vect{\bar{N}}),\vect{S})$. Here, $\vect{\bar{N}} \in \mathbb{C}^{N_r \times B}$ is the mean disturbance and $\vect{S} \in \mathbb{C}^{N_r B \times N_r B }$ is the positive definite covariance matrix. The additive disturbance term describes the receiver noise and the interference from all other concurrent transmissions, which might involve the same or other receivers. The latter is commonly referred to as \emph{pilot contamination} in the large-scale MIMO literature \cite{Marzetta2010a,Jose2011b,Hoydis2013a,Rusek2013a,Larsson2014a} and can in general have a non-zero line-of-sight component. The analysis herein holds for any $\vect{\bar{N}}$ and $\vect{S}$, but some typical special cases are described and evaluated numerically in Section \ref{section:Numerical_Evaluation}.

Vectorizing the received matrix in \eqref{eq_pilotmodel} yields
\begin{equation*}
\vect{y} = \widetilde{\vect{P}} \vect{h}  + \vect{n}
\end{equation*}
where $\vect{y} = \vectorize(\vect{Y}), \widetilde{\vect{P}} \triangleq (\vect{P}^T \! \kron \, \vect{I}), \vect{h} = \vectorize(\vect{H}) $ and $\vect{n} = \vectorize(\vect{N})$. This transforms the matrix estimation in \eqref{eq_pilotmodel} into the canonical form of vector estimation in \cite{Kay1993a} which enables the use of classical estimation results.

If the channel and disturbance statistics (i.e., $\vect{\bar{H}},\vect{R},\vect{\bar{N}}$ and $\vect{S}$) are perfectly known at the receiver, the Bayesian MMSE estimator of the MIMO channel is \cite{Kay1993a,Kotecha2004a,Liu2007a,Bjornson2010a}
\begin{equation} \label{eq_MMSE_channel_matrix}
\begin{split}
\MMSE{\vect{h}} = \vectorize(\MMSE{\vect{H}}) = \vect{\bar{h}} + \vect{R} \widetilde{\vect{P}}^H
\left(\widetilde{\vect{P}} \vect{R} \widetilde{\vect{P}}^H +
\vect{S} \right)^{-1}  \vect{d}
\end{split}
\end{equation}
where $\vect{\bar{h}} = \vectorize(\vect{\bar{H}}), \vect{\bar{n}} = \vectorize(\vect{\bar{N}})$ and $\vect{d} = \vect{y} - \widetilde{\vect{P}} \vect{\bar{h}} - \vect{\bar{n}} $. We measure the performance in terms of the estimation MSE. Using the MMSE estimator, it follows that
\begin{equation}\label{eq:MSE_MMSE}
\textrm{MSE} =\mathbb{E}\{ \|\vect{H} - \MMSE{\vect{H}} \|_F^2 \} = \tr \left( ( \vect{R}^{-1} + \widetilde{\vect{P}}^H
\vect{S}^{-1} \widetilde{\vect{P}}
 )^{-1} \right).
\end{equation}

Alternatively, if the channel distribution is unknown to the receiver, the classic MVU estimator is \cite[Chapter 4]{Kay1993a}
\begin{equation} \label{eq_MVU_channel_matrix}
\begin{split}
\MVU{\vect{h}} = \vectorize(\MVU{\vect{H}}) = \left(\widetilde{\vect{P}}^H \vect{S}^{-1} \widetilde{\vect{P}} \right)^{-1} \widetilde{\vect{P}}^H \vect{S}^{-1} (\vect{y - \bar{n}}).
\end{split}
\end{equation}
The corresponding performance measure is then the estimation variance $\mathbb{E}\{ \|\vect{H} - \MVU{\vect{H}} \|_F^2 \} = \tr \left( ( \widetilde{\vect{P}}^H
\vect{S}^{-1} \widetilde{\vect{P}} )^{-1} \right)$.

Note that the mean matrices of the channel and the disturbance have no impact on the performance with MMSE and MVU estimation.
Moreover,
\begin{equation}
\tr \left( ( \vect{R}^{-1} + \widetilde{\vect{P}}^H
\vect{S}^{-1} \widetilde{\vect{P}}
 )^{-1} \right) < \tr \left( ( \widetilde{\vect{P}}^H
\vect{S}^{-1} \widetilde{\vect{P}} )^{-1} \right)
\end{equation}
for any $\vect{R} \neq \vect{0}$, thus the MMSE estimator achieves a better average estimation performance than the MVU estimator since it utilizes the channel statistics.

\begin{remark}[Arbitrary Statistics] \label{remark:arbitrary-statistics}
While having Gaussian channels and disturbance is a well-accepted assumption in conventional MIMO systems, the channel modeling for large-scale MIMO is still in its infancy. By increasing the number of antennas we improve the spatial resolution of the array which eventually may invalidate the rich-scattering assumption that is behind the use of Gaussian channel distributions \cite{Rusek2013a}. However, we stress that the results of this paper can be applied and give reasonable performance under any arbitrary statistical distributions on the channel and disturbance; this is since \eqref{eq_MMSE_channel_matrix} is also the linear MMSE estimator and \eqref{eq_MVU_channel_matrix} is the best linear unbiased estimator (BLUE) in cases when only the first two moments of $\vect{H}$ and/or $\vect{N}$ are known \cite{Kay1993a,Bjornson2010a}.
\end{remark}

Recall that we assumed that the statistical parameters $\vect{\bar{H}},\vect{R},\vect{\bar{N}}$, and $\vect{S}$ of the channel and disturbance are known at the receiver. Since user mobility and large-scale fading cause continuous changes in the statistics, this implicitly means that the receiver can keep track of these changes. Such tracking can, for example, be achieved by exploiting the pilot signals on multiple flat-fading subcarriers since the large-scale fading properties can be transformed between different adjacent subcarriers \cite{Aste1998a,Chalise2004a}. Interestingly, the coherence time of the long-term statistics is relatively short; the measurements in \cite{Viering2002a} observe coherence times of $5$--$23$ seconds, depending on the propagation environment. High user velocity or rapid scheduling decisions in neighboring systems can further reduce the coherence time. More importantly, the number of channel realizations within each coherence time of the statistics is around $13$--$126$, according to \cite{Viering2002a}. This means that the matrix inversion in the MMSE estimator has to be recomputed frequently.

\subsection{Complexity Issues in Large-Scale MIMO Systems}
\label{subsec:complexity-issues}

The main computational complexity when computing the MMSE and MVU estimators in \eqref{eq_MMSE_channel_matrix} and \eqref{eq_MVU_channel_matrix} lies in solving a linear system of equations or, equivalently, in computing the matrix inversions directly. Both approaches have computational complexities that scale as $\mathcal{O}(M^3)$, where $M \triangleq B N_r$ is the matrix dimension.\footnote{Note that $\mathcal{O}(M^3)$ refers to the complexity scaling of the classical inversion algorithms, such as Gaussian elimination and inversion based on Cholesky decomposition \cite{Boyd2008a}. The exponent is reduced to $\mathcal{O}(M^{2.8074})$ by Strassen's algorithm in \cite{Strassen1969a}, which is a divide-an-conquer algorithm that exploits that $2 \times 2$ matrices can be multiplied efficiently. Using the complexity expressions in \cite{Strassen1969a}, it is easy to show that the algorithm is only computationally beneficial for very large matrices (e.g., $M \gtrsim 8000$) due to heavy overhead computations. It also has other drawbacks, such as lower computational accuracy and that the matrix dimensions must be $M = 2^{k}$ for some integer $k$. The exponent can be further reduced to $\mathcal{O}(M^{2.373})$ \cite{Williams2012a}, but at the cost of more overhead that pushes the breaking point to even higher values of $M$. In this paper, we propose new estimators with the complexity scaling $\mathcal{O}(M^{2})$, which both is a asymptotically better and is proved to be beneficial at large but practical $M$.} This complexity is relatively modest in conventional MIMO communication systems where $2 \times 2$, $4 \times 4$, or $8 \times 8$ are typical configurations.

Recently, there is an increasing interest in large-scale MIMO systems where there might be hundreds of antennas at one side of the link \cite{Marzetta2010a,Jose2011b,Hoydis2013a,Rusek2013a,Larsson2014a,Baldemair2013a}. To excite all channel dimensions, the pilot length $B$ should be of the same order as $N_t$. Large-scale MIMO systems are therefore envisioned to operate in TDD mode and exploit channel reciprocity to always have $N_t < N_r$ in the channel estimation phase---$N_r$ can even be orders of magnitude larger than $N_t$ without degrading the estimation performance \emph{per} antenna element.

Observe that in a potential future large-scale MIMO system with $N_r=200$ and $N_t=B=20$, the MMSE and MVU estimators would require inverting matrices of size $4000 \times 4000$ (or similarly, solving a linear system of equations with $4000$ unknown variables) which has a complexity at the order of $3.4 \cdot 10^{11}$ floating-point operations, see Section \ref{subsec:Summ-Comp-Comp} for details. This massive matrix manipulation needs to be redone every few seconds since $\vect{R}$ and $\vect{S}$ change due to mobility. Motivated by these facts, the purpose of this paper is to develop alternative channel estimators that allow for balancing between computational/hardware complexity and estimation performance.

\subsection{A Diagonalization Approach to Complexity Reduction}
\label{subsec:naive}

There is a special case when the computational complexity of MMSE estimation can be greatly reduced, namely when the matrices $\vect{R}$, $\vect{S}$, and $\vect{P}$ are all diagonal matrices. The matrix $\widetilde{\vect{P}} \vect{R} \widetilde{\vect{P}}^H +
\vect{S}$ is then also diagonal which allows for computing $(\widetilde{\vect{P}} \vect{R} \widetilde{\vect{P}}^H +
\vect{S} )^{-1}$ by simply inverting each diagonal element. The corresponding complexity is only $8M - 1 = \mathcal{O}(M)$ FLOPs.
This special case is, unfortunately, of limited practical interest for large-scale MIMO systems which are prone to non-negligible spatial channel correlation and pilot contamination.\footnote{The elements of each column of $\vect{H}$ are highly correlated due the insufficient antenna spacing and limited richness of the scattering around the large array at the receiver. The correlation between the columns depends more on the scattering and size of the small array at the transmitter, thus the correlation might be weaker but complete independence is seldom seen in practice. In the ideal case of exactly independent columns, the covariance matrix $\widetilde{\vect{P}} \vect{R} \widetilde{\vect{P}}^H + \vect{S}$ is block-diagonal which can be exploited for complexity reduction. The complexity scaling of the  MMSE estimation is, however, still cubic in $N_r$ and the proposed estimators have a computational advantage when $N_r$ is sufficiently large; see Section \ref{subsec:Summ-Comp-Comp}.}

Inspired by this special case, a simple approach to complexity reduction is to diagonalize the covariance matrices $\vect{R}$ and $\vect{S}$ by replacing all off-diagonal elements by zero. Let $\vect{R}_{\diag}$ and $\vect{S}_{\diag}$ denote the corresponding matrices, assume $B= N_t$, and set $\vect{P} = \sqrt{\mathcal{P}_t} \vect{I}$ where $\mathcal{P}_t$ is the average pilot power. The MMSE estimator in \eqref{eq_MMSE_channel_matrix} is approximated as
\begin{equation} \label{eq:diagonalized-estimator}
\widehat{\tilde{\vect{h}}} = \vect{\bar{h}} + \sqrt{\mathcal{P}_t}  \vect{R}_{\diag}
\left( \mathcal{P}_t \vect{R}_{\diag}  + \vect{S}_{\diag} \right)^{-1}  \vect{d}
\end{equation}
where the matrix $\vect{R}_{\diag} \left( \vect{R}_{\diag}  + \vect{S}_{\diag} \right)^{-1}$ can be precomputed with a computational complexity proportional to $M$. From now on, we refer to \eqref{eq:diagonalized-estimator} as the \textit{diagonalized estimator}. It achieves the following MSE.

\begin{theorem}\label{theorem:MSE_diagonalized}
The diagonalized estimator in \eqref{eq:diagonalized-estimator} with $\vect{P} = \sqrt{\mathcal{P}_t} \vect{I}$ achieves the MSE
\begin{equation} \label{eq:MSE-diagonalized}
\tr \left( \left( \vect{R}_{\diag}^{-1} + \mathcal{P}_t \vect{S}_{\diag}^{-1} \right)^{-1} \right).
\end{equation}

In noise-limited scenarios with $\vect{S} = \sigma^2 \vect{I}$, the MSE of the diagonalized estimator goes to zero as the power $\mathcal{P}_t \rightarrow \infty$.
\end{theorem}
\begin{IEEEproof}
The diagonalized estimator in \eqref{eq:diagonalized-estimator} estimates each channel element separately, thus the MSE is equivalent to that of MMSE estimation with $\vect{R}_{\diag}$ as channel covariance matrix and $\vect{S}_{\diag}$ as disturbance covariance matrix \cite{Bjornson2010a}. This gives the MSE expression in \eqref{eq:MSE-diagonalized}. By letting $\mathcal{P}_t \rightarrow \infty$ in  \eqref{eq:MSE-diagonalized}, it follows directly that the MSE approaches zero asymptotically.
\end{IEEEproof}

This theorem shows that the diagonalized estimator performs well in noise-limited scenarios with high signal-to-noise ratio (SNR). Unfortunately, the simulations in Section \ref{section:Numerical_Evaluation} reveals that this is the only operating regime where it is comparable to the MMSE estimator. More precisely, the drawback of the diagonalized estimator is that it does not exploit the statistical dependence neither between the received pilot signals nor between the channel coefficients. We recall from \cite{Bjornson2010a} that exploiting such dependence (e.g., spatial correlation) can give great MSE improvements. Therefore, the next section develops a new sophisticated type of channel estimators that reduces the computational complexity of MMSE estimation while retaining the full statistical information. These estimators are great complements to the diagonalized estimator since they perform particularly well at low to medium SNRs and under interference.

\section{Low-Complexity Bayesian PEACH Estimators}
\label{section:Low_complexity estimators}

In this section, we propose several low-complexity Bayesian channel estimators based on the concept of polynomial expansion. To understand the main idea, we first state the following lemma which is easily proved by using standard Taylor series.

\begin{lemma} \label{lemma:inversion-expansion}
For any Hermitian matrix $\vect{X} \in \mathbb{C}^{N \times N}$, with bounded eigenvalues $| \lambda_n(\vect{X})|<1$ for all $n$, it holds that
\begin{equation} \label{eq:inversion-expansion}
\left( \vect{I} - \vect{X} \right)^{-1} = \sum_{l=0}^{\infty} \vect{X}^{l}.
\end{equation}
\end{lemma}

Observe that the impact of $\vect{X}^{l}$ in \eqref{eq:inversion-expansion} reduces with $l$, as $\lambda_n(\vect{X})^{l}$ for each eigenvalue. It therefore makes sense to consider $L$-degree polynomial expansions of the matrix inverse using only the terms $l=0,\ldots,L$. In principle, the inverse of each eigenvalue is then approximated by an $L$-degree Taylor polynomial, thus $L$ needs \emph{not} to scale with the matrix dimension to achieve a certain accuracy per element. Instead, $L$ can be selected to balance between low approximation error and low complexity. To verify this independency in the area of estimation, we investigate the MSE performance of large-scale MIMO systems of different dimensions in Section \ref{section:Numerical_Evaluation}. We observe an almost identical performance for a fixed $L$ when we vary the number of antennas. Note that a similar remark was made in \cite{Honig2001a} where the authors show that their system performance metric does not depend on the system dimensions but only the filter rank.

In order to apply Lemma \ref{lemma:inversion-expansion} on matrices with any eigenvalue structure, we obtain the next result
which is similar to \cite{Sessler2005a}.

\begin{proposition} \label{prop:inversion-expansion}
For any positive-definite Hermitian matrix $\vect{X}$ and any $0 < \alpha < \frac{2}{\max_n \lambda_n(\vect{X})}$, it holds that
\begin{equation}
\vect{X}^{-1} = \alpha \big( \vect{I} - (\vect{I} - \alpha \vect{X}) \big)^{-1} = \alpha \sum_{l=0}^{L} (\vect{I} - \alpha \vect{X})^l + \vect{E}
\end{equation}
where $\alpha \sum_{l=0}^{L} (\vect{I} - \alpha \vect{X})^l$ is an $L$-degree polynomial approximation and the error term $\vect{E}$ is bounded as $\| \vect{E}\|_2 = \mathcal{O} \big( \|(\vect{I} - \alpha \vect{X})\|_2^{L+1} \big)$. The error vanishes as $L \rightarrow \infty$.
\end{proposition}

\subsection{Unweighted PEACH Estimator}
\label{subsec:PEACH}

Applying the approximation in Proposition \ref{prop:inversion-expansion} on the MMSE estimator in \eqref{eq_MMSE_channel_matrix} gives the low-complexity $L$-degree \emph{Polynomial ExpAnsion CHannel (PEACH)} estimator which we denote by $\PEACH{\vect{h}} = \vectorize(\PEACH{\vect{H}})$ and define as
\begin{equation} \label{eq_PEACH_estimator}
\begin{split}
\PEACH{\vect{h}} \triangleq  \vect{\bar{h}} +  \vect{R} \widetilde{\vect{P}}^H
\sum_{l=0}^{L}  \alpha \big(\vect{I} - \alpha (\widetilde{\vect{P}} \vect{R} \widetilde{\vect{P}}^H + \vect{S} ) \big)^{l}  \vect{d}.
\end{split}
\end{equation}
Note that \eqref{eq_PEACH_estimator} does not involve any inversions. Furthermore, the polynomial structure $\sum_{l=0}^{L} \vect{X}^l \vect{d}$ lends itself to a recursive computation
\begin{equation} \label{eq_recursive_computation}
\sum_{l=0}^{L} \vect{X}^l \vect{d} = \vect{d} + \vect{X} \bigg( \vect{d} + \vect{X} \Big( \vect{d} + \vect{X} \big(\vect{d} + \vect{X} ( \ldots ) \big) \Big) \bigg)
\end{equation}
where $\vect{X} = \vect{I} - \alpha (\widetilde{\vect{P}} \vect{R} \widetilde{\vect{P}}^H + \vect{S} )$ for the PEACH estimator. The key property of \eqref{eq_recursive_computation} is that it only involves matrix-vector multiplications, which have a complexity of $\mathcal{O}(M^2)$ instead of the cubic complexity of matrix-matrix multiplications \cite{Boyd2008a}. The computational complexity of \eqref{eq_PEACH_estimator} is therefore $\mathcal{O}(L M^2)$ where $M \triangleq B N_r$. Whenever $L \ll M$, $\mathcal{O}(L M^2)$ is a large complexity reduction as compared to $\mathcal{O}(M^3)$ for the original MMSE estimator. Furthermore, the recursive structure enables an efficient multistage hardware implementation similar to the detection implementation illustrated in \cite[Fig.~1]{Moshavi1996a}.

\begin{theorem}\label{theorem:MSE_PEACH}
The PEACH estimator in \eqref{eq_PEACH_estimator} achieves the MSE
\begin{equation}\label{eq:MSE_PEACH}
\tr \left(\vect{R} +  \vect{R} \widetilde{\vect{P}}^H  \vect{A}_L (\widetilde{\vect{P}} \vect{R} \widetilde{\vect{P}}^H + \vect{S} )  \vect{A}_L^H  \widetilde{\vect{P}} \vect{R} - 2  \vect{R} \widetilde{\vect{P}}^H  \vect{A}_L  \widetilde{\vect{P}} \vect{R} \right)
\end{equation}
where $\vect{A}_L = \sum_{l=0}^{L}  \alpha \big(\vect{I} - \alpha (\widetilde{\vect{P}} \vect{R} \widetilde{\vect{P}}^H + \vect{S} ) \big)^{l}$.
\end{theorem}
\begin{IEEEproof}
This theorem follows from direct computation of the MSE using the definition $\mathrm{MSE}  = \mathbb{E}\{ \| \vect{h} - \PEACH{\vect{h}} \|^2 \}$.
\end{IEEEproof}

It remains to select the scaling parameter $\alpha$ to satisfy the convergence condition in Proposition \ref{prop:inversion-expansion}. From a pure complexity point of view, we can select $\alpha$ to be equal to $\frac{2}{\tr(\widetilde{\vect{P}} \vect{R} \widetilde{\vect{P}}^H + \vect{S})}$  \cite{Lei1998a}. However, the choice of $\alpha$ also determines the convergence speed of the polynomial expansion. Among the values that satisfy the condition in Proposition \ref{prop:inversion-expansion}, the choice
\begin{equation}\label{eq:alpha_PEACH}
\begin{split}
\alpha = \frac{2}{\max_n \lambda_n(\widetilde{\vect{P}} \vect{R} \widetilde{\vect{P}}^H + \vect{S}) + \min_n \lambda_n(\widetilde{\vect{P}} \vect{R} \widetilde{\vect{P}}^H + \vect{S}) }
\end{split}
\end{equation}
minimizes the spectral radius of $\big(\vect{I} - \alpha (\widetilde{\vect{P}} \vect{R} \widetilde{\vect{P}}^H + \vect{S} ) \big)$ and therefore provides the fastest asymptotic convergence speed  \cite{Sessler2005a}.\footnote{The error term in Proposition \ref{prop:inversion-expansion} is bounded by $\mathcal{O} \big( \|(\vect{I} - \alpha \vect{X})\|_2^{L+1} \big)$. The spectral norm is minimized by making the largest and smallest eigenvalues symmetric around the origin \cite{Sessler2005a}: $\max_n \lambda_n(\vect{I} - \alpha \vect{X}) = - \min_n \lambda_n(\vect{I} - \alpha \vect{X})$. By solving for $\alpha$ we obtain $\alpha = 2/ ( \max_n \lambda_n( \vect{X}) + \min_n \lambda_n( \vect{X}))$ which becomes \eqref{eq:alpha_PEACH} for the problem at hand.} Although the computation of the extreme eigenvalues is generally quite expensive, these eigenvalues can be approximated with lower complexity. For example, as mentioned earlier, if the convergence speed is not the main concern $\max_n \lambda_n(\widetilde{\vect{P}} \vect{R} \widetilde{\vect{P}}^H + \vect{S}) + \min_n \lambda_n(\widetilde{\vect{P}} \vect{R} \widetilde{\vect{P}}^H + \vect{S})$ simply can be estimated by $\tr(\widetilde{\vect{P}} \vect{R} \widetilde{\vect{P}}^H + \vect{S})$. Alternatively, the smallest eigenvalue can be taken as the noise variance and largest eigenvalue can be approximated using some upper bound on the pilot power and on the average channel attenuation to the receiver. In general, a low-complexity method to approximate the extreme eigenvalues of any arbitrary covariance matrix was proposed in \cite{Sessler2005a}, based on the Gershgorin circle theorem \cite{Bhatia1997a}. This approach exploits the structure of the matrix imposed by the system setup to improve the convergence speed. For more details on how to choose $\alpha$ with low-complexity and compute the extreme eigenvalues we refer to \cite{Sessler2005a}.

\subsection{Weighted PEACH Estimator}

Although the PEACH estimator \eqref{eq_PEACH_estimator} converges to the MMSE estimator as $L \rightarrow \infty$, it is generally not the best $L$-degree polynomial estimator at any finite $L$. More specifically, instead of multiplying each term in the sum with $\alpha$, we can assign different weights and optimize these for the specific degree $L$.
In this way, we obtain the \emph{weighted PEACH estimator} which we denote as $\WPEACH{\vect{h}} = \vectorize(\WPEACH{\vect{H}})$ and define as
\begin{equation} \label{eq_weighted_PEACH_estimator}
\begin{split}
\WPEACH{\vect{h}}  \triangleq \vect{\bar{h}} + \vect{R} \widetilde{\vect{P}}^H
\sum_{l=0}^{L} w_l \alpha_{\mathrm{w}}^{l+1} \big(\widetilde{\vect{P}} \vect{R} \widetilde{\vect{P}}^H + \vect{S}  \big)^{l}  \vect{d}
\end{split}
\end{equation}
where $\vect{w} = [w_0,\ldots,w_L]^T$ are scalar weighting coefficients.\footnote{W-PEACH is obtained by expanding each $(\vect{I} - \alpha (\widetilde{\vect{P}} \vect{R} \widetilde{\vect{P}}^H + \vect{S} ) )^{l}$ as a binomial series, collecting terms, and replacing constant factors with weights.} Observe that the $\alpha$-parameter, now denoted $\alpha_{\mathrm{w}}$, is redundant and can be set to one. For numerical reasons, it might still be good to select
\begin{equation}
\alpha_{\mathrm{w}} \leq \frac{1}{\max_n \lambda_n(\widetilde{\vect{P}} \vect{R} \widetilde{\vect{P}}^H + \vect{S})}
\end{equation}
since this makes all the eigenvalues of $\alpha_{\mathrm{w}}^{l+1} \big(\widetilde{\vect{P}} \vect{R} \widetilde{\vect{P}}^H + \vect{S}  \big)^{l}$ smaller than one and thus prevent them from growing unboundedly as $l$ becomes large. This simplifies the implementation of the following theorem, which finds the weighting coefficients that minimize the MSE.

\begin{theorem} \label{theorem:MSE-minimizing-weights}
The MSE $\mathbb{E}\{ \| \vect{h} - \WPEACH{\vect{h}} \|^2 \}$ is minimized by
\begin{equation} \label{eq_optimal_coefficients}
\vect{w}_{\mathrm{opt}} = [w_0^{\mathrm{opt}} \, \ldots \, w_L^{\mathrm{opt}}]^T = \vect{A}^{-1} \vect{b}
\end{equation}
where the $ij$th element of $\vect{A} \in \mathbb{C}^{L+1 \times L+1}$ and the $i$th element of $\vect{b} \in \mathbb{C}^{L+1}$ are
\begin{equation} \label{eq_A_b_matrices}
\begin{split}
[\vect{A}]_{ij} &= \alpha_{\mathrm{w}}^{i+j}  \tr\left(  \vect{R} \widetilde{\vect{P}}^H (\widetilde{\vect{P}} \vect{R} \widetilde{\vect{P}}^H + \vect{S})^{i+j-1}     \widetilde{\vect{P}} \vect{R}  \right), \\
[\vect{b}]_i &= \alpha_{\mathrm{w}}^{i} \tr\left( \vect{R} \widetilde{\vect{P}}^H (\widetilde{\vect{P}} \vect{R} \widetilde{\vect{P}}^H + \vect{S})^{i-1}  \widetilde{\vect{P}} \vect{R} \right).
\end{split}
\end{equation}
The resulting MSE of the W-PEACH estimator is
\begin{equation}\label{eq:minimum MSE}
\mathrm{MSE}  = \tr(\vect{R}) - \vect{b}^H \vect{A}^{-1} \vect{b}.
\end{equation}
\end{theorem}
\begin{IEEEproof}
The W-PEACH estimator achieves an MSE of
\begin{equation} \label{eq:proof-MSE-expression}
\begin{split}
&\mathrm{MSE}  = \mathbb{E}\{ \| \vectorize(\vect{H}) - \vectorize(\WPEACH{\vect{H}}) \|_F^2 \} \\
&= \tr\Bigg( \vect{R}  - \vect{R} \widetilde{\vect{P}}^H
\sum_{l=0}^{L} (w_l+w_l^*) \alpha_{\mathrm{w}}^{l+1} \vect{Z}^{l}  \widetilde{\vect{P}} \vect{R} \\
&\quad +
\sum_{l_1=0}^{L} \sum_{l_2=0}^{L} w_{l_1} w^*_{l_2} \alpha_{\mathrm{w}}^{l_1+l_2+2} \vect{R} \widetilde{\vect{P}}^H \vect{Z}^{l_1+l_2+1}     \widetilde{\vect{P}} \vect{R}  \Bigg)
\end{split}
\end{equation}
where $\vect{Z} = \widetilde{\vect{P}} \vect{R} \widetilde{\vect{P}}^H + \vect{S}$.
For a given pilot matrix $\vect{P}$ and polynomial degree $L$, the coefficients $w_0,\ldots,w_L$ can be selected to minimize the MSE as
\begin{equation}
\minimize{w_0,\ldots,w_L} \quad \mathrm{MSE}.
\end{equation}
The solution to this unconstrained optimization problem is achieved by computing the partial derivatives with respect to each coefficient and looking for stationary points:
\begin{equation} \label{eq:derivative-MSE}
\begin{split}
&\frac{\partial }{\partial w_l} \mathrm{MSE}  =  - \alpha_{\mathrm{w}}^{l+1} \tr\left( \vect{R} \widetilde{\vect{P}}^H \vect{Z}^{l}  \widetilde{\vect{P}} \vect{R} \right) \\
&+ \sum_{l_2=0}^{L} w^*_{l_2} \tr\left(  \vect{R} \widetilde{\vect{P}}^H \alpha_{\mathrm{w}}^{l_1+l_2+2} \vect{Z}^{l_1+l_2+1}    \widetilde{\vect{P}} \vect{R}  \right).
\end{split}
\end{equation}
By equating to zero for each $l=0,\ldots,L$, we achieve $L+1$ linear equations that involve the $L+1$ unknown coefficients. These are $\vect{A} \vect{w} = \vect{b}$ with $\vect{A}, \vect{b}$ as in \eqref{eq_A_b_matrices}; note that we made a change of variables $i=l_1+1$ and $j=l_2+1$ for $\vect{A}$ and $i=l+1$ for $\vect{b}$, because the sums in \eqref{eq:derivative-MSE} begin at 0 while the indices of matrices/vectors usually begin at 1. The MSE minimizing weights are now computed as in \eqref{eq_optimal_coefficients}.

Finally, we note that, using $\vect{A}, \vect{b}$ in \eqref{eq_A_b_matrices}, the MSE expression in \eqref{eq:proof-MSE-expression} can be expressed as $\tr(\vect{R}) + \vect{w}^H \vect{A} \vect{w}  - \vect{b}^H \vect{w}  - \vect{w}^H \vect{b}$. For optimal weights $\vect{w}_{\mathrm{opt}} = \vect{A}^{-1} \vect{b}$, the minimum MSE becomes \eqref{eq:minimum MSE}.
\end{IEEEproof}

Observe that the MSE expressions of PEACH and W-PEACH in \eqref{eq:MSE_PEACH} and \eqref{eq:minimum MSE}, respectively, are independent of the mean matrices of the channel and the disturbance. Therefore, the performance is the same as in our conference paper \cite{Shariati2013a}, where we assumed zero-mean channel and disturbance.

From \eqref{eq:proof-MSE-expression} in the proof of Theorem \ref{theorem:MSE-minimizing-weights}, we also obtain the MSE expression
\begin{equation}\label{eq:general MSE-expression}
\mathrm{MSE}(\vect{w})  = \tr(\vect{R}) + \vect{w}^H \vect{A} \vect{w}  - \vect{b}^H \vect{w}  - \vect{w}^H \vect{b}
\end{equation}
for the W-PEACH estimator with any choice of the weighting coefficients.

\begin{remark}[Weights of the PEACH estimator] The PEACH estimator can also be expressed as a W-PEACH estimator using certain weights. To find these weights, we observe that
\begin{equation*}
\begin{split}
\sum_{l=0}^{L} & \alpha \big(\vect{I} - \alpha (\widetilde{\vect{P}} \vect{R} \widetilde{\vect{P}}^H + \vect{S} ) \big)^{l} \\
&=\sum_{l=0}^{L} \alpha \sum_{n=0}^{l} \binom{l}{n} (- \alpha)^n (\widetilde{\vect{P}} \vect{R} \widetilde{\vect{P}}^H + \vect{S} )^{n} \vect{I}^{l-n} \\
&=\sum_{l=0}^{L}  \sum_{n=0}^{l} \binom{l}{n} (-1)^n \alpha^{n+1} (\widetilde{\vect{P}} \vect{R} \widetilde{\vect{P}}^H + \vect{S} )^{n}.
\end{split}
\end{equation*}
By gathering all terms that belong to a certain exponent $n$, we see that
\begin{equation}
w_n = (-1)^n  \sum_{l=n}^{L} \binom{l}{n}.
\end{equation}
Plugging these weights into \eqref{eq:general MSE-expression} yields an alternative way of computing the MSE of the PEACH estimator.
\end{remark}

Although Theorem \ref{theorem:MSE-minimizing-weights} provides the optimal weights, the computational complexity is $\mathcal{O}(M^3)$ since it involves pure matrix multiplications of the form $\vect{Z}^{i}$. This means that computing the optimal weights for the W-PEACH estimator has the same asymptotic complexity scaling as computing the conventional MMSE estimator. To benefit from the weight optimization we thus need to find an approximate low-complexity approach to compute the weights, which is done in the next subsection. Note that the weights cannot be optimized by random matrix theory (as was done for multiuser detection in \cite{Muller2001a,Hoydis2011d} and precoding in \cite{Zarei2013a,Mueller2014a,Kammoun2014b}) due to lack of randomness in the MMSE estimation expression in \eqref{eq_MMSE_channel_matrix}.

\begin{remark}[Low-Complexity Classical PEACH Estimators]
 Following the same approach as used to derive low-complexity PEACH estimators for the Bayesian case, we form the corresponding low-complexity estimators to approximate the classic MVU estimator in \eqref{eq_MVU_channel_matrix}. Note that if the quality of the channel covariance matrix estimate is very poor, then the MVU estimator performs better than the MMSE estimator.

First, we define a regularization factor $\epsilon > 0$ which in the form of $\epsilon \vect{I}$ is added to $(\widetilde{\vect{P}}^H \vect{S}^{-1} \widetilde{\vect{P}})$. Then, we use the matrix inversion lemma which results in
 \begin{equation} \label{eq_MVU_regularized}
\begin{aligned}
\MVU{\vect{h}}^{\epsilon} &= \left( \epsilon \vect{I} + \widetilde{\vect{P}}^H \vect{S}^{-1} \widetilde{\vect{P}} \right)^{-1} \widetilde{\vect{P}}^H \vect{S}^{-1} (\vect{y - \bar{n}}) & \\
&=  \widetilde{\vect{P}}^H  \left( \widetilde{\vect{P}} \widetilde{\vect{P}}^H +  \epsilon\vect{S} \right)^{-1}  (\vect{y - \bar{n}})  = \MVU{\vect{h}} \mid_{\epsilon \rightarrow 0}.&
\end{aligned}
\end{equation}
 The approximation in Proposition \ref{prop:inversion-expansion} can now be applied. The set of low-complexity PEACH estimators obtained by this approach are
\begin{equation}\label{eq:MVU_PEACH_estimator}
\PEACH{\vect{h}}^{\mathrm{MVU}}=   \widetilde{\vect{P}}  \sum_{l=0}^L  \alpha \left( \vect{I} - \alpha (\widetilde{\vect{P}} \widetilde{\vect{P}}^H + \epsilon \vect{S})^l \right) (\vect{y - \bar{n}})
\end{equation}
and
\begin{equation}\label{eq:MVU_WPEACH_estimator}
\WPEACH{\vect{h}}^{\mathrm{MVU}}=   \widetilde{\vect{P}}  \sum_{l=0}^L  w_l \alpha_{\mathrm{w}}^{l+1}   (\widetilde{\vect{P}} \widetilde{\vect{P}}^H + \epsilon \vect{S})^l  (\vect{y - \bar{n}}).
\end{equation}
Observe that the last equality in \eqref{eq_MVU_regularized} equals to \eqref{eq_MMSE_channel_matrix} if $ \vect{R} = \frac{1}{\epsilon} \vect{I}$, therefore all the results presented in Theorems \ref{theorem:MSE_PEACH}  and \ref{theorem:MSE-minimizing-weights} can be derived for $\PEACH{\vect{h}}^{\mathrm{MVU}}$ and $\WPEACH{\vect{h}}^{\mathrm{MVU}}$ in a similar way.
\end{remark}

\begin{remark}[Other PEACH estimators]
The PE technique can be applied to any type of channel estimators that involve matrix inversions. For example, \cite{Eldar2004a} derives a robust estimator, the \emph{minimax regret estimator}, under certain uncertainty and statistical assumptions. This estimator has a similar expression as the MMSE estimator, but involves other matrices. Hence, the PE technique is straightforward to apply and the weights can be optimized similar to what is described herein.
\end{remark}

\subsection{Low-Complexity Weights}\label{subsection:Low-Complexity Weights}

Next, we propose a low-complexity algorithm to compute weights for the W-PEACH estimator. We exploit that
\begin{equation}
(\widetilde{\vect{P}} \vect{R} \widetilde{\vect{P}}^H + \vect{S}) = \mathbb{E}\{ \vectorize(\vect{Y}) \vectorize(\vect{Y})^H \} = \lim_{T \rightarrow \infty} \frac{1}{T} \sum_{t=1}^{T} \vect{y}_t \vect{y}_t^H
\end{equation}
where $\vect{y}_t = \vectorize(\vect{Y})$ denotes the received signal at estimation time instant $t$. This means that $(\widetilde{\vect{P}} \vect{R} \widetilde{\vect{P}}^H + \vect{S})$ is closely approximated by the sample covariance matrix $\frac{1}{T} \sum_{t=1}^{T} \vect{y}_t \vect{y}_t^H$ if the number of samples $T$ is large. Although one generally needs $T \gg B N_r$ to get a consistent approximation, we can get away with much smaller $T$ since we only use it to compute traces---this is verified numerically in Section \ref{section:Numerical_Evaluation}.

For any fixed $T \geq 1$ and $i\geq 1$, we now observe that
\begin{align} \label{eq:low-complex1}
&\tr\left(  \vect{R} \widetilde{\vect{P}}^H (\widetilde{\vect{P}} \vect{R} \widetilde{\vect{P}}^H + \vect{S})^{i}     \widetilde{\vect{P}} \vect{R}  \right) \\
&\approx \tr\left(  \vect{R} \widetilde{\vect{P}}^H (\widetilde{\vect{P}} \vect{R} \widetilde{\vect{P}}^H + \vect{S})^{i-1} \left(\frac{1}{T} \sum_{t=1}^{T}  \vect{y}_t \vect{y}_t^H \right)    \widetilde{\vect{P}} \vect{R}  \right) \\
&= \frac{1}{T} \sum_{t=1}^{T} \vect{y}_t^H \left( \widetilde{\vect{P}} \vect{R}^2 \widetilde{\vect{P}}^H (\widetilde{\vect{P}} \vect{R} \widetilde{\vect{P}}^H + \vect{S})^{i-1} \right) \vect{y}_t. \label{eq:low-complex3}
\end{align}

Since the elements of $\vect{A}$ and $\vect{b}$ in \eqref{eq_A_b_matrices} are of the form in \eqref{eq:low-complex1}, we can approximate each element using \eqref{eq:low-complex3}.\footnote{Note that $b_0 = \tr( \widetilde{\vect{P}} \vect{R}^2 \widetilde{\vect{P}}^H )$ needs to be treated differently since there is no $(\widetilde{\vect{P}} \vect{R} \widetilde{\vect{P}}^H + \vect{S})$ term. In the case when
$\widetilde{\vect{P}}^H \widetilde{\vect{P}}$ is a scaled identity matrix, we only need to compute $\tr( \vect{R}^2  )$ which can be done efficiently since only the diagonal elements of $\vect{R}^2$ are of interest. Otherwise, one can select a set of $T$ vectors $\vect{v}_i \! \sim \! \mathcal{CN}(\vect{0},\vect{I})$ and apply the approximation $\tr( \widetilde{\vect{P}} \vect{R}^2 \widetilde{\vect{P}}^H ) \approx \frac{\alpha_{\mathrm{w}}}{T} \sum_{i=1}^{T} \vect{v}_i^H \widetilde{\vect{P}} \vect{R}^2 \widetilde{\vect{P}}^H \vect{v}_i$. This is the approach included in Algorithm \ref{algorithm_low-complexity}.} By computing/updating these approximations over a sliding time window of length $T$, we obtain Algorithm \ref{algorithm_low-complexity}. At any time instant $t$, this algorithm computes approximations of $\vect{A},\vect{b}$, denoted by $\widetilde{\vect{A}}_{t},\tilde{\vect{b}}_{t}$, by using the received signals $\vect{y}_t,\ldots,\vect{y}_{t-T+1}$.
These are used to compute approximate weights $\vect{w}_{\mathrm{approx},t}$. To reduce the amount of computations, $\widetilde{\vect{A}}_{t},\tilde{\vect{b}}_{t}$ are obtained from $\widetilde{\vect{A}}_{t-1},\tilde{\vect{b}}_{t-1}$ by adding one term per element based on the current received signal $\vect{y}_{t}$ and removing the impact of the old received signal $\vect{y}_{t-T}$ (which is now outside the time window). The algorithm can be initialized in any way; for example, by accumulating $T$ received signals to fill the time window.

\begin{algorithm}[t]
  \KwIn{Polynomial degree $L$ and time window $T$\;}
  \KwIn{Current time $t$\;}
  \KwIn{New and old received signals $\vect{y}_t,\vect{y}_{t-T}$\;}
  \KwIn{Approximations $\widetilde{\vect{A}}_{t-1},\tilde{\vect{b}}_{t-1}$ at previous time $t\!-\!1$\;}
  Set $[\widetilde{\vect{A}}_{t}]_{ij} = [\widetilde{\vect{A}}_{t-1}]_{ij} $ \begin{displaymath}
  \begin{split}
  &+ \frac{ \alpha_{\mathrm{w}}^{i+j}}{T} \vect{y}_t^H \left( \widetilde{\vect{P}} \vect{R}^2 \widetilde{\vect{P}}^H (\widetilde{\vect{P}} \vect{R} \widetilde{\vect{P}}^H + \vect{S})^{i+j-2} \right) \vect{y}_t \\
  &- \frac{ \alpha_{\mathrm{w}}^{i+j}}{T} \vect{y}_{t-T}^H \left( \widetilde{\vect{P}} \vect{R}^2 \widetilde{\vect{P}}^H (\widetilde{\vect{P}} \vect{R} \widetilde{\vect{P}}^H + \vect{S})^{i+j-2} \right) \vect{y}_{t-T} \,\,\, \forall i,j\;
  \end{split}
  \end{displaymath}

  Set $[\tilde{\vect{b}}_{t}]_i = [\tilde{\vect{b}}_{t-1}]_i$ \begin{displaymath}
  \begin{split}
  & + \frac{ \alpha_{\mathrm{w}}^{i}}{T} \vect{y}_t^H \left( \widetilde{\vect{P}} \vect{R}^2 \widetilde{\vect{P}}^H (\widetilde{\vect{P}} \vect{R} \widetilde{\vect{P}}^H + \vect{S})^{i-2} \right) \vect{y}_t \\& - \frac{ \alpha_{\mathrm{w}}^{i}}{T} \vect{y}_{t-T}^H \left( \widetilde{\vect{P}} \vect{R}^2 \widetilde{\vect{P}}^H (\widetilde{\vect{P}} \vect{R} \widetilde{\vect{P}}^H + \vect{S})^{i-2} \right) \vect{y}_{t-T} \,\,\,\, \forall i \geq 2\;
     \end{split}
  \end{displaymath}

  Set $[\tilde{\vect{b}}_{t}]_1 = \frac{\alpha_{\mathrm{w}}}{T} \sum_{i=1}^{T} \vect{v}_i^H \widetilde{\vect{P}} \vect{R}^2 \widetilde{\vect{P}}^H \vect{v}_i$ for $\vect{v}_i \! \sim \! \mathcal{CN}(\vect{0},\vect{I})$\;

  Compute $\vect{w}_{\mathrm{approx},t} = \widetilde{\vect{A}}_{t}^{-1} \tilde{\vect{b}}_{t}$\;
  \KwOut{Approximate weights $\vect{w}_{\mathrm{approx},t}$ at time $t$\;}
\caption{Low-complexity weights for W-PEACH} \label{algorithm_low-complexity}
\end{algorithm}

The asymptotic complexity of computing the elements in $\widetilde{\vect{A}}_{t}$ and $\tilde{\vect{b}}_{t}$ is $\mathcal{O}(L M^2)$ \emph{FLOPs per time instant}. For each element, we need to compute  a series of multiplications between vectors and matrices of complexity $\mathcal{O}(M^2)$. This is explained in detail in Section \ref{subsec:Summ-Comp-Comp} where we derive the exact computational complexity. Next, $\vect{w}_{\mathrm{approx},t}$ is obtained by solving an $L$-dimensional system of equations, which has complexity $\mathcal{O}(L^3)$. Finally, the W-PEACH estimate is computed in the recursive manner described in Section \ref{subsec:PEACH} with a complexity of $\mathcal{O}(L M^2)$. To summarize, the W-PEACH estimator along with Algorithm \ref{algorithm_low-complexity} has a computational complexity of $\mathcal{O}(L M^2 + L^3)$.

One additional feature of Algorithm \ref{algorithm_low-complexity} is that it can easily be extended to practical scenarios where only imperfect estimates of the covariance matrices $\vect{R}$ and $\vect{S}$ are available. Apart from enabling adaptive tracking of the slow variations in the channel and disturbance statistics, this practical scenario is relevant to understand how sensitive Bayesian channel estimators are to mismatches in the statistical knowledge. We perform a numerical study in Section \ref{section:Numerical_Evaluation}, based on the statistical estimation described in the next subsection.

\subsection{Imperfect Covariance Matrix Estimation}
\label{subsec:sample-covariance}

Suppose we want to obtain some covariance matrix $\vect{C}$ from $N$ observations $\vect{c}_1,\ldots,\vect{c}_N$, where $\vect{C}$ might be $\vect{R}$ or $\widetilde{\vect{P}} \vect{R} \widetilde{\vect{P}}^H + \vect{S}$.
The sample covariance matrix $\vect{C}_{\mathrm{sample}} \triangleq \frac{1}{N} \sum_{i=1}^{N} \vect{c}_i \vect{c}_i^H$ is conventionally used to estimate $\vect{C}$. However, this approach is unsuitable for large-scale systems where it can be hard to accumulate more samples than the dimension of $\vect{C}$, which is $N_t N_r$ for the channel covariance matrix $\vect{R}$. In fact, the sample covariance matrix is not even invertible if the number of samples is smaller than the matrix dimension. Instead of using the pure sample covariance matrix, we suggest to follow a similar approach as in \cite{Ledoit2004a} and use a new estimator $\hat{\vect{C}}$ which is an affine function of the sample covariance matrix $\vect{C}_{\mathrm{sample}}$. In \cite{Ledoit2004a}, the authors have shown that this estimator is a better fit for large-dimensional covariance matrices.

Here, different from the diagonal loading approach in \cite{Ledoit2004a}, where they consider an affine combination of the identity matrix and the sample covariance matrix, we assume $\hat{\vect{C}} = \kappa \vect{C}_{d} + (1 - \kappa) \vect{C}_{\mathrm{sample}}$ where $\vect{C}_{d}$ is the diagonal matrix comprising the diagonal elements of $\vect{C}_{\mathrm{sample}}$ and $\kappa$ is chosen to minimize the squared difference $\mathbb{E}\{\|\hat{\vect{C}} - \vect{C}\|_F^2\}$. The advantage of $\hat{\vect{C}}$ is that the diagonal elements converge quickly with $N$ to their true values, while the reliance on the off-diagonal elements is controlled by the parameter $\kappa$. The optimal $\kappa$ is given by the following theorem.

\begin{theorem} \label{theorem:kappa-opt}
The solution $\kappa^\star$ to the optimization problem $\underset{\kappa}{\mathrm{min}} \hspace{.2cm} \mathbb{E}\{\|\hat{\vect{C}} - \vect{C}\|^2\}$, where $\hat{\vect{C}} = \kappa \vect{C}_{d} + (1 - \kappa) \vect{C}_{\mathrm{sample}}$, is
\begin{equation}\label{eq:kappa_opt}
\kappa^\star = \frac{\Phi(\vect{C}_{\mathrm{sample}}) - \frac{1}{2} \Psi(\vect{C}_d,\vect{C}_{\mathrm{sample}})}{\Phi(\vect{C}_{\mathrm{sample}}) + \Phi(\vect{C}_d) - \Psi(\vect{C}_d,\vect{C}_{\mathrm{sample}})}
\end{equation}
where $\Phi(\vect{C}_{\mathrm{sample}}) = \mathbb{E}\{\|\vect{C}_{\mathrm{sample}} - \vect{C}\|_F^2\}$, $\Phi(\vect{C}_d) = \mathbb{E}\{\|\vect{C}_d - \vect{C}\|_F^2\}$ and $\Psi(\vect{C}_d,\vect{C}_{\mathrm{sample}}) = \mathbb{E}\{ \tr \big((\vect{C}_d - \vect{C})(\vect{C}_{\mathrm{sample}} - \vect{C}) \big)\}$.
\end{theorem}
\begin{IEEEproof}
The objective function can be rewritten as
\begin{equation*}
\begin{split}
&\mathbb{E}\{\| \kappa \vect{C}_d + (1 - \kappa) \vect{C}_{\mathrm{sample}} - \vect{C} - \kappa\vect{C} + \kappa\vect{C} \|_F^2\} \\
& = \mathbb{E}\{\| \kappa (\vect{C}_d - \vect{C}) \|_F^2\}  + \mathbb{E}\{\| (1-\kappa) (\vect{C}_{\mathrm{sample}} - \vect{C}) \|_F^2\} \\
&+ 2 \kappa (1 - \kappa) \mathbb{E}\{ \tr \big((\vect{C}_d - \vect{C})(\vect{C}_{\mathrm{sample}} - \vect{C}) \big)\}.
\end{split}
\end{equation*}
Considering $\Phi(\vect{C}_{\mathrm{sample}})$, $\Phi(\vect{C}_d)$, and $\Psi(\vect{C}_d,\vect{C}_{\mathrm{sample}})$, the first-order optimality condition is
\begin{equation*}
2 \kappa \Phi(\vect{C}_d) - 2 (1 - \kappa) \Phi(\vect{C}_{\mathrm{sample}}) + (1 - 2 \kappa) \Psi(\vect{C}_d,\vect{C}_{\mathrm{sample}}) = 0,
\end{equation*}
which yields the optimal solution $\kappa^\star$ in \eqref{eq:kappa_opt}.
\end{IEEEproof}

Note that as the number of samples $N$ grows large, the optimal $\kappa^\star$ will be smaller which implies that we put larger trust in the sample covariance matrix. In Section \ref{section:Numerical_Evaluation}, we apply this theory to the channel covariance matrix and compare the estimation performance when using $\hat{\vect{R}}$ to performance with the true covariance matrix $\vect{R}$. Interestingly, we observe that the proposed W-PEACH estimator adapts itself very well to imperfect statistics.

\subsection{Asymptotic and Exact Computational Complexity}\label{subsec:Summ-Comp-Comp}

The asymptotic complexity of the conventional estimators, the diagonalized estimator described in Section \ref{subsec:naive}, and the proposed PEACH estimators are summarized as follows:

\vskip+3mm

\begin{center}
     \begin{tabular}{ | c | c |}
     \hline
     Channel Estimators & Computational Complexity \\ \hline
     MMSE and MVU & $\mathcal{O}(B^3 N_r^3)$ \\ \hline
     Diagonalized & $\mathcal{O}(B N_r)$ \\ \hline
     PEACH & $\mathcal{O}(L B^2 N_r^2)$ \\ \hline
     W-PEACH & $\mathcal{O}(L B^2 N_r^2 + L^3)$ \\ \hline
     \end{tabular}
 \end{center}

\vskip+2mm

These asymptotic complexity numbers are supported by an exact complexity analysis below. We note that the cubic complexity scaling in $B N_r$ for the conventional MMSE and MVU estimators is reduced to linear complexity in the diagonalized approach and squared complexity for the proposed PEACH estimators. The degree $L$ of the polynomial expansion has a clear impact on the complexity, but recall that it needs not scale with $B N_r$ \cite{Honig2001a}. This property is illustrated in the next section, where we also show that small values on $L$ yields good performance.

The high complexity of the conventional estimators is not an issue if the channel and disturbance statistics are fixed over a very long time horizon; the system can then simply compute the inverse and then use it over and over again. As described in Section \ref{subsec:complexity-issues}, the statistics change continuously in practice and it is thus necessary to redo the inversion every few seconds.\footnote{The MMSE estimator can be implemented recursively \cite{Matz2005a}, which is suitable for tracking variations in the covariance matrices. The complexity of each recursion is $\mathcal{O}(M^2)$, but we need more than $M$ recursions (per long-term statistics coherence time) to obtain a stable covariance estimate \cite{Matz2005a}. Hence, the recursive implementation also has a cubic complexity.} To make a precise and fair comparison, we need to consider the relationship between the coherence time of the long-term statistics, $\tau_s$, and the channel coherence time, denoted by $\tau_c$. The analysis below reveals how the computational complexity, in terms of the number of FLOPs, depends on the system dimensions, polynomial degree $L$, and the coherence times $\tau_s$ and $\tau_c$. For the sake of brevity, we consider complex-valued FLOPs and neglect the computational small complexity of scalar multiplications and additions of matrices and vectors.

The ratio $Q = \frac{\tau_s}{\tau_c}$ describes how stationary the channel statistics are \cite{Viering2002a}, in terms of how many channel realizations that fit into the coherence time of the statistics. The propagation environment has significant impact on this ratio; for example, in \cite{Viering2002a} the authors have shown that $Q$ equals $13$, $108$ and $126$ for indoor, rural and urban environments, respectively, under their measurement setup. Smaller number are expected when the transmitter/receiver travel with high velocity. Similarly, the disturbance statistics can change rapidly if it contains interference from other systems (particularly if adaptive scheduling is performed) \cite{Yang2011a}. For a given total time $T_{\mathrm{tot}}$, the computational complexity for each of the estimators consists of two parts: one part which can be precomputed once per coherence time of the statistics (i.e., $k_s=\frac{T_{\mathrm{tot}}}{\tau_s}$ times) and one part that is computed at channel realization (i.e., $k_c=\frac{T_{\mathrm{tot}}}{\tau_c}$). Note that $k_c = Q k_s$.

We use the notation $M = N_r B$ and $N = N_r N_t$. For given vectors $\vect{x, y} \in \mathbb{C}^{N \times 1}$ and matrices $\vect{A} \in \mathbb{C}^{M \times N}$ and $\vect{B}\in \mathbb{C}^{N \times P}$, there are $MP(2N - 1)$, $M(2N - 1)$ and $2N - 1$ FLOPs required for the matrix-matrix product $\vect{AB}$, matrix-vector product $\vect{Ax}$, and vector-vector product $\vect{x}^H \vect{y}$, respectively. In the special case of $M = P$ and $\vect{C} = \vect{AB}$ being symmetric, only $\frac{1}{2}M(M + 1)(2N - 1)$ FLOPs are required to obtain $\vect{C}$. Moreover, the Cholesky factorization of a positive definite matrix $\vect{A} \in \mathbb{C}^{M \times M}$ is computed using $\frac{1}{3}M^3$ FLOPs. To solve a linear system of equations $\vect{Ax} = \vect{b}$, where $\vect{b} \in \mathbb{C}^{M \times 1}$, by exploiting Cholesky factorization and back-substitution, a total of $\frac{1}{3}M^3 + 2M^2$ FLOPs is needed \cite{Boyd2008a}.

We denote the total computational complexity in FLOPs by $\chi$. For the MMSE estimator, the two parts $\mathbf{U}_{\mathrm{MMSE}} = \vect{R} \widetilde{\vect{P}}^H
\left(\widetilde{\vect{P}} \vect{R} \widetilde{\vect{P}}^H +
\vect{S} \right)^{-1}$ and $\mathbf{v}= \widetilde{\vect{P}}^H \vect{\bar{h}} + \vect{\bar{n}}$ are computed once per $\tau_s$ and the parts  $\mathbf{d = y - v}$ and $\vect{\bar{h}} + \mathbf{Ud} $ once per $\tau_c$. It results in a total computational complexity of $\chi_{\mathrm{MMSE}} = k_c \big[ N(2M-1)\big] + k_s \big[ \frac{1}{3}M^3 + (3N - 0.5)M^2 + (2N^2 + 2N - \frac{3}{2})M \big]$ in FLOPs.

For the MVU estimator, there is $\mathbf{U}_{\mathrm{MVU}} = \left(\widetilde{\vect{P}}^H \vect{S}^{-1} \widetilde{\vect{P}} \right)^{-1} \widetilde{\vect{P}}^H \vect{S}^{-1}$ which is computed once per $\tau_s$, and the parts $\vect{y - \bar{n}}$ (neglected) and $\mathbf{U}_{\mathrm{MVU}} (\vect{y - \bar{n}}) $ computed once per $\tau_c$, yielding to $\chi_{\mathrm{MVU}} = k_c \big[ N(2M-1)\big] + k_s \big[ \frac{1}{3}M^3 + 2NM^2 + (3N^2 + N)M + \frac{1}{3}N^3 - 0.5N^2 - 0.5N \big]$.

For the proposed PEACH and W-PEACH estimators, only $\mathbf{v}$ is computed once per  $\tau_s$. The rest of the computations take place once per $\tau_c$. As described in \eqref{eq_recursive_computation}, the polynomial $\sum_{l=0}^{L} \vect{X}^l \vect{d}$, where $\vect{X} = \vect{I} - \alpha (\widetilde{\vect{P}} \vect{R} \widetilde{\vect{P}}^H + \vect{S} )$, is computed recursively. The first term $\vect{d}$ is readily available. The second term $\vect{Xd}$ is computed as a series of matrix-vector products. First, we compute $\vect{S}\vect{d}$ and $\widetilde{\vect{P}}^H \vect{d}$. Next, we multiply $\vect{R}$ with the resulting vector of $(\widetilde{\vect{P}}^H \vect{d})$, and then $\widetilde{\vect{P}}$ is multiplied with the vector $(\vect{R} \widetilde{\vect{P}}^H  \vect{d})$. The vector $\vect{d} - \alpha \widetilde{\vect{P}} \vect{R} \widetilde{\vect{P}}^H \vect{d} - \alpha \vect{S d}$ is then computed. We repeat this procedure $L$ times and exploit $\vect{X}^{l} \vect{d}$ to compute $\vect{X}^{l+1} \vect{d}$. For the PEACH estimator, the total computational complexity is $\chi_{\mathrm{PEACH}} = k_c \big[ 2LM^2 + ((4L+2)N - 2L)M + 2(L+1)N^2 - 2(L+1)N \big] + k_s \big[ M(2N-1)\big]$ FLOPs.

The polynomial structure of W-PEACH estimator requires the same number of FLOPs as the PEACH estimator, but there are two additional sources of computations: solving the linear system of equations $\vect{A^{-1} b}$ to compute the weight vector $\vect{w}_{\mathrm{opt}}$ (which requires $\frac{1}{3}(L+1)^3 + 2(L+1)^2$ FLOPs) and using Algorithm \ref{algorithm_low-complexity} to find the approximated elements of $\vect{A}$ and $\vect{b}$. The computational complexity of Algorithm \ref{algorithm_low-complexity} is counted by considering the following: Firstly, we only need to obtain the elements in $\tilde{\vect{A}_t}$, since all the elements of $\tilde{\vect{b}_t}$ can be extracted out from $\tilde{\vect{A}_t}$. In particular, all the elements contain similar terms $\vect{Z}^k$ with $\vect{Z} = \tilde{\vect{P}} \vect{R} \tilde{\vect{P}}^H + \vect{S}$, where $0 \leq k \leq 2L$ in $\tilde{\vect{A}_t}$ and $0 \leq k \leq L-1$ in $\tilde{\vect{b}_t}$. Secondly, we exploit the fact that $\vect{Z}^k \vect{y_t}$ for $0 \leq k \leq L$ has been already computed in the estimator expression $\sum_{l=0}^L \vect{Z}^l \vect{y_t}$. Thirdly, to determine all the elements in $\tilde{\vect{A}_t}$, we first need to compute $\vect{Z}^k \vect{y_t}$ for $L+1 \leq k \leq 2L$ which results in doing a recursive matrix-vector multiplication $L$ times (i.e., $L[M(2M-1) + N(2M-1) + N(2N-1) + M(2N -1)]$ FLOPs) and then compute $\vect{y_t}^H \tilde{\vect{P}} \vect{R}^2 \tilde{\vect{P}}^H$. Note that this term can be considered as the multiplication of $\vect{y_t}^H \tilde{\vect{P}} \vect{R}$ and $\vect{R} \tilde{\vect{P}}^H$, where the first term $\vect{y_t}^H \tilde{\vect{P}} \vect{R}$ has already been computed. This results in two matrix-vector products (i.e., $N(2N-1) + M(2N -1)$ FLOPs). Finally, for each element, we have the vector-vector multiplication $(\vect{y_t}^H \tilde{\vect{P}} \vect{R}^2 \tilde{\vect{P}}^H) (\vect{Z}^k \vect{y_t})$ resulting in $(2L+1)(2M -1)$ FLOPs. To summarize, for the W-PEACH estimator, we have $\chi_{\mathrm{W-PEACH}} = k_c \big[ 4LM^2 + (8L + 4)MN + (4L + 4)N^2 + M - (4L + 3)N + \frac{1}{3}L^3 + 3L^2 + 3L + \frac{4}{3} \big] + k_s \big[ M(2N-1) \big]$ FLOPs.

In the following table we summarize the exact total computational complexity of the different estimators when $B = N_t$, which makes $M=N$.

\begin{center}
     \begin{tabular}{ | c | c |}
     \hline
     Estimators &  FLOPs \\ \hline
     MMSE & $k_c \big[ 2M^2 \!-\! M \big] \!+\! k_s \big[ \frac{16}{3}M^3 \!+\! \frac{3}{2}M^2 \!-\! \frac{3}{2}M \big]$ \\ \hline
     MVU & $k_c \big[ 2M^2 \!-\! M \big] \!+\! k_s \big[ \frac{17}{3}M^3 \!+\! \frac{1}{2}M^2 \!-\! \frac{1}{2}M \big]$ \\ \hline
     PEACH & $k_c \big[ (8L \!+\! 4)M^2 \!-\! (4L \!+\! 2)M \big] \!+\! k_s \big[  2M^2 \!-\! M \big]$ \\ \hline
     W-PEACH & $k_c \big[ (16L \!+\! 8)M^2 \!-\! (4L \!+\! 2)M $  \\ \hline
     & $\!+\! \frac{1}{3}L^3 \!+\! 3L^2 \!+\! 3L \!+\! \frac{4}{3} \big] \!+\! k_s \big[ 2M^2 \!-\! M  \big]$ \\ \hline
     \end{tabular}
 \end{center}

Now, recalling $k_c = Q k_s$ and comparing the dominating terms of the MMSE and PEACH estimators, we can obtain a condition (the relation between the values $L$, $Q$ and $M$) for when the PEACH estimators are less complex than the MMSE estimator. This condition is
\begin{equation} \label{eq:complexity-condition-PEACH}
\frac{16}{3}M \geq 8 Q L + 2 Q \Rightarrow M \geq Q \left( \frac{3}{2}L + \frac{3}{8} \right)
\end{equation}
for the PEACH estimator, and
\begin{equation} \label{eq:complexity-condition-WPEACH}
\frac{16}{3}M \geq 16 Q L + 6 Q \Rightarrow M \geq Q \left( 3L + \frac{9}{8} \right)
\end{equation}
for the W-PEACH estimator. This implies that only under certain numbers of the channel stationarity, polynomial degree, and the number of antennas, PEACH estimators are less complex than the MMSE estimator and will provide reasonable performance. For the practical values of $Q=50$ and $L=2$, \eqref{eq:complexity-condition-PEACH} and \eqref{eq:complexity-condition-WPEACH} show that the PEACH and W-PEACH estimators outperform the MMSE estimator in terms of complexity for $M = N_tN_r \geq 167$ and $M \geq 357$, respectively. Hence, the PEACH estimator is practically useful for setup such as $N_t=2$ and $N_r=100$ or $N_t=1$ and $N_r=200$, similarly the W-PEACH estimator for $N_t=4$ and $N_r=100$ or $N_t=1$ and $N_r=400$.

As demonstrated by the complexity analysis, the PEACH estimators are computed using only matrix-vector multiplications. This is a standard operation that can easily be parallelized and implemented using efficient integrated circuits. On the contrary, the matrix inversions in the MMSE/MVU estimators are known to be complicated to implement in hardware \cite{Shepard2012a}.  Consequently, whenever the PEACH estimators and MMSE/MVU estimators are similar in terms of FLOPs, the computational delays and energy consumption are probably lower when implementing the proposed PEACH estimators.

\section{Performance Evaluation}
\label{section:Numerical_Evaluation}

In this section, we analyze and illustrate the performance of the proposed diagonalized, PEACH, and W-PEACH estimators. The analysis so far has been generic with respect to the disturbance covariance matrix $\vect{S}$. Here, we consider two scenarios: noise-limited and cellular networks with pilot contamination. We describe the latter scenario in more detail since it is one of the main challenges in the development of large-scale MIMO systems \cite{Rusek2013a}. This section provides asymptotic analysis and numerical results for both scenarios.

\subsection{Noise-Limited Scenario}
\label{subsection:noise-limited}

A commonly studied scenario is when there is only uncorrelated receiver noise; thus $\vect{S} = \sigma^2 \vect{I}$ where $\sigma^2$ is the noise variance. As the pilot power grows large, the MSE of the MMSE estimator is known to go asymptotically to zero \cite{Kay1993a,Kotecha2004a,Liu2007a,Bjornson2010a}. We proved in Theorem~\ref{theorem:MSE_diagonalized} that the diagonalized estimator has the same asymptotically optimal behavior in the high-power regime. Here, in the following proposition, we derive the asymptotic behavior of the PEACH and W-PEACH estimators in the noise-limited scenario.

\begin{proposition}\label{proposition:Noise-limited}
As the pilot power $\mathcal{P}_t \rightarrow \infty$ with the pilot matrix $\vect{P} = \sqrt{\mathcal{P}_t} \vect{I}$, the MSEs of the PEACH and W-PEACH estimators converge to the non-zero MSE floors
\begin{equation}\label{eq:Noise_limited_PEACH}
\tr \left(\vect{R} +  \vect{R}  \vect{B}_L \vect{R} \vect{B}_L^H \vect{R} - 2  \vect{R}  \vect{B}_L \vect{R} \right)
\end{equation}
and
\begin{equation}\label{eq:Noise_limited_WPEACH}
\tr\left( \vect{R} - \vect{\tilde{b}}^H \vect{\tilde{A}}^{-1} \vect{\tilde{b}}\right)
\end{equation}
respectively, where
$\Lambda = \max_n \lambda_n( \vect{R} ) + \min_n \lambda_n( \vect{R} )$, $\vect{B}_L = \frac{2}{\Lambda} \sum_{l=0}^{L} \big( \vect{I} -  \frac{2}{\Lambda}  \vect{R}  \big)^{l}$,  $[\vect{\tilde{A}}]_{ij} = \alpha_{\mathrm{w}}^{i+j}  \tr \left( (\vect{R})^{i+j+1}\right)$, and $[\vect{\tilde{b}}]_i = \alpha_{\mathrm{w}}^i  \tr \left( (\vect{R})^{i+1} \right)$.
\end{proposition}
\begin{IEEEproof}
First, we focus on the PEACH estimator with $\vect{P} = \sqrt{\mathcal{P}_t} \vect{I}$, where the MSE expression in \eqref{eq:MSE_PEACH} can be rewritten as
\begin{equation}\label{eq:approx_MSE_PEACH}
\tr \left(\vect{R} +   \vect{R}  (\mathcal{P}_t \vect{A}_L) \big(\vect{R} + \frac{1}{\mathcal{P}_t} \vect{S} \big) (\mathcal{P}_t \vect{A}_L^H)  \vect{R} - 2  \vect{R} (\mathcal{P}_t  \vect{A}_L)  \vect{R} \right).
\end{equation}
Observe that $\mathcal{P}_t \vect{A}_L =  \sum_{l=0}^{L} \mathcal{P}_t \alpha \big(\vect{I} - \mathcal{P}_t \alpha ( \vect{R} + \frac{1}{\mathcal{P}_t} \vect{S} ) \big)^{l} = \vect{B}_L$ as $\mathcal{P}_t \rightarrow \infty$, because $\frac{1}{\mathcal{P}_t} \vect{S} = \frac{\sigma^2}{\mathcal{P}_t} \vect{I} \rightarrow \vect{0}$ and $\mathcal{P}_t \alpha = \frac{2}{ \max_n \lambda_n( \vect{R} + \frac{1}{\mathcal{P}_t} \vect{S} ) + \min_n \lambda_n( \vect{R} + \frac{1}{\mathcal{P}_t} \vect{S}) } \rightarrow \frac{2}{\Lambda}$ using the expression of $\alpha$ in \eqref{eq:alpha_PEACH}.\footnote{Similar MSE floors for the PEACH estimator are obtained for any way of selecting $\alpha$, as a function of $\mathcal{P}_t$, to satisfy the condition in Proposition~\ref{prop:inversion-expansion}.} By taking the limit $\mathcal{P}_t \rightarrow \infty$ in the MSE expression \eqref{eq:approx_MSE_PEACH} and exploiting the aforementioned limits $\mathcal{P}_t \vect{A}_L \rightarrow \vect{B}_L$ and $\frac{1}{\mathcal{P}_t} \vect{S} \rightarrow \vect{0}$ we obtain the non-zero MSE floor \eqref{eq:Noise_limited_PEACH} which is independent of $\mathcal{P}_t$.

Next, for the W-PEACH estimator, the minimum MSE is $\vect{b}^H \vect{A}^{-1} \vect{b}$ where $\vect{A}$ and $\vect{b}$ are given in Theorem \ref{theorem:MSE-minimizing-weights}. For normalization reasons we define $\vect{D} \triangleq \mathrm{diag}(1, \frac{1}{\mathcal{P}_t}, \frac{1}{\mathcal{P}_t^2}, \ldots , \frac{1}{\mathcal{P}_t^L})$ and note that $\vect{b}^H \vect{A}^{-1} \vect{b} = (\vect{D} \vect{b})^H (\vect{D}\vect{A}\vect{D}) ^{-1} (\vect{D} \vect{b}) $. The limit, as $\mathcal{P}_t \rightarrow \infty$, of each element of $\vect{D}\vect{A}\vect{D}$ and $\vect{D} \vect{b}$ are
\begin{equation}
\begin{split}
 [\vect{D}\vect{A}\vect{D}]_{ij} &= \frac{1}{\mathcal{P}_t^{i+j}} \alpha_{\mathrm{w}}^{i+j}  \tr\left(  \vect{R} \widetilde{\vect{P}}^H (\widetilde{\vect{P}} \vect{R} \widetilde{\vect{P}}^H + \vect{S})^{i+j-1}     \widetilde{\vect{P}} \vect{R}  \right) \\
 &= \alpha_{\mathrm{w}}^{i+j}  \tr\left( \vect{R} (  \vect{R} + \frac{\sigma^2}{\mathcal{P}_t} \vect{I})^{i+j-1} \vect{R} \right) \\
 &\rightarrow \alpha_{\mathrm{w}}^{i+j}  \tr \left( (\vect{R})^{i+j+1}\right)
\end{split}
\end{equation}
and
\begin{equation}
\begin{split}
 [\vect{D} \vect{b}]_i &= \frac{1}{\mathcal{P}_t^{i}} \alpha_{\mathrm{w}}^{i} \tr\left( \vect{R} \widetilde{\vect{P}}^H (\widetilde{\vect{P}} \vect{R} \widetilde{\vect{P}}^H + \vect{S})^{i-1}  \widetilde{\vect{P}} \vect{R} \right)\\
&= \alpha_{\mathrm{w}}^{i} \tr\left( \vect{R} ( \vect{R}  +  \frac{\sigma^2}{\mathcal{P}_t} \vect{I})^{i-1}  \vect{R} \right)\\
&\rightarrow \alpha_{\mathrm{w}}^i  \tr \left( (\vect{R})^{i+1}\right),
\end{split}
\end{equation}
under the condition that $\alpha_{\mathrm{w}}$ is fixed (recall that for the W-PEACH estimator $\alpha_{\mathrm{w}}$ can be selected arbitrarily).
By denoting the limits of $\vect{D}\vect{A}\vect{D}$ and $\vect{D} \vect{b}$ as $\vect{\tilde{A}}$ and $\vect{\tilde{b}}$, respectively,
the MSE expression \eqref{eq:minimum MSE} converges to the non-zero floor
\begin{equation*}
 \tr\left( \vect{R} - \vect{\tilde{b}}^H \vect{\tilde{A}}^{-1} \vect{\tilde{b}}\right).
\end{equation*}
This MSE floor is independent of $\mathcal{P}_t$ and is only a function of channel covariance matrix and its moments. However, by similar justification as that of used for the PEACH estimator (i.e., having  $\alpha \propto \mathcal{P}_t^{-1}$) we observe that the MSE expression \eqref{eq:minimum MSE} converges to a non-zero error floor independent of $\mathcal{P}_t$.
\end{IEEEproof}

This proposition shows that the MSEs of the PEACH and W-PEACH estimators exhibit non-zero error floors as the power increases.
This reveals that, in order to reduce complexity, it is better to ignore the spatial channel correlation (as with the diagonalized estimator) than approximating the full matrix inversion (as with the PEACH estimators) in the high-power regime of noise-limited scenarios.

\subsection{Pilot Contamination Scenario}
\label{subsection:pilot-contamination}

A scenario that has received much attention in the large-scale MIMO literature is when there is
disturbance from simultaneous reuse of pilot signals in neighboring cells \cite{Marzetta2010a,Jose2011b,Hoydis2013a,Rusek2013a,Larsson2014a,Yin2013a,Mueller2013a}. Such reuse is often necessary due to the finite channel coherence time (i.e., the time that a channel estimate can be deemed accurate), but leads to a special form of interference called pilot contamination. It can be modeled as\footnote{Cell $i$ can use an arbitrary pilot matrix $\vect{P}_i$, but only pilot matrices with overlapping span (i.e., $\vect{P}_i \vect{P}^H \neq \vect{0}$) cause interference to the desired pilot signaling. Therefore, the case of a common reused pilot matrix $\vect{P}_i =\vect{P} \,\, \forall i \in \mathcal{I}$ is the canonical example, while extensions to partially overlapping pilots are achieved by removing the non-overlapping parts (e.g., by considering $\vect{Y} \vect{P}^H$ as the effective received signal). Moreover, it is assumed in \eqref{eq:interference-with-contamination} that the interfering pilots are synchronized with the desired pilot and that the delays between cells are negligible. These are, essentially, worst-case assumptions and alternative unsynchronized scenarios have recently been analyzed in \cite{Fernandes2013a}.}
\begin{equation} \label{eq:interference-with-contamination}
\vect{N} = \sum_{i \in \mathcal{I}} \vect{H}_i \vect{P} + \widetilde{\vect{N}}
\end{equation}
where $\mathcal{I}$ is the set of interfering cells, $\vect{H}_i$ is the channel from the transmitter in the $i$th interfering cell to the receiver in the cell under study, and $\vectorize(\widetilde{\vect{N}}) \sim \mathcal{CN}(\vect{0},\sigma^2 \vect{I})$ is the uncorrelated receiver noise. If $\vect{H}_i$ is Rayleigh fading with $\vectorize(\vect{H}_i) \sim \mathcal{CN}(\vect{0},\boldsymbol{\Sigma}_i)$, then
\begin{equation} \label{eq:pilot-contamination-covariance}
\vect{S} =  \sum_{i \in \mathcal{I}} \widetilde{\vect{P}} \boldsymbol{\Sigma}_i \widetilde{\vect{P}}^H + \sigma^2 \vect{I}.
\end{equation}
Note that only the sum covariance matrix $\sum_{i \in \mathcal{I}}  \boldsymbol{\Sigma}_i$ needs to be known when computing the proposed PEACH estimators. Moreover, only the diagonal elements of the sum covariance matrix are used by the diagonalized estimator.

When \eqref{eq:pilot-contamination-covariance} is substituted into the PEACH and W-PEACH estimator expressions in \eqref{eq_PEACH_estimator} and \eqref{eq_weighted_PEACH_estimator} we get contaminated disturbance terms of the form $\vect{R} \widetilde{\vect{P}}^H \widetilde{\vect{P}} \boldsymbol{\Sigma}_i \widetilde{\vect{P}}^H$. These terms are small if $\vect{R}$ and $\boldsymbol{\Sigma}_i $ have very different span, or if $\tr(\boldsymbol{\Sigma}_i)$ is weak altogether---this is easily observed if $\widetilde{\vect{P}}^H \widetilde{\vect{P}}$ is a scaled identity matrix. Similar observations were recently made in the capacity analysis of \cite{Hoydis2013a} and when developing a pilot allocation algorithm in \cite{Yin2013a}. Under certain conditions, the subspaces of the useful channel and pilot contamination can be made orthogonal by coordinated allocation of pilot resources across cells \cite{Yin2013a} or by exploiting both received pilot and data signals for channel estimation as in \cite{Mueller2013a}.

Similar to the noise-limited scenario, we want to understand how the MSE with different estimators behave as the pilot power $\mathcal{P}_t \rightarrow \infty$. We begin with the MMSE estimator and the proposed diagonalized estimator, for which the MSEs saturates in the asymptotic regime under pilot contamination.

\begin{proposition}\label{proposition:Pilot contamination MSE floor}
As the pilot power $\mathcal{P}_t \rightarrow \infty$ with the pilot matrix $\vect{P} = \sqrt{\mathcal{P}_t} \vect{I}$, the MSEs with the MMSE estimator and diagonalized estimator converge to the MSE floors
\begin{equation}\label{eq:MMSE error floor}
\tr\left( \vect{R} - \vect{R}^2 (\vect{R} + \sum_{i \in \mathcal{I}}  \boldsymbol{\Sigma}_i)^{-1} \right)
\end{equation}
and
\begin{equation}\label{eq:diagonalized error floor}
\sum_{j=1}^{N_tN_r} r_j - \sum_{j=1}^{N_tN_r} \frac{ r_j^2 }{ r_j + \sum_{i \in \mathcal{I}} \sigma_{i,j} },
\end{equation}
respectively, where $r_j$ and $\sigma_{i,j}$ are the $j\textit{th}$ elements of $\vect{R}_{\mathrm{diag}}$ and $\boldsymbol{\Sigma}_{\mathrm{diag},i}$, respectively. Note that $\vect{S}_{\mathrm{diag}} = \mathcal{P}_t\sum_{i \in \mathcal{I}} \boldsymbol{\Sigma}_{\mathrm{diag},i} + \sigma^2 \vect{I}$.
\end{proposition}
\begin{IEEEproof}
We start by noting that the MSE of the MMSE estimator behaves as
\begin{equation*}
\begin{split}
\mathrm{MSE} &= \tr \left( \vect{R} - \vect{R} \widetilde{\vect{P}}^H (\widetilde{\vect{P}} \vect{R} \widetilde{\vect{P}}^H + \vect{S})^{-1} \widetilde{\vect{P}} \vect{R} \right)\\
&= \tr\left( \vect{R} -  \vect{R}^2 \left[ (\vect{R} + \sum_{i \in \mathcal{I}}  \boldsymbol{\Sigma}_i) + \frac{\sigma^2}{\mathcal{P}_t} \vect{I} \right]^{-1} \right)\\
& \rightarrow \tr\left( \vect{R} - \vect{R}^2 (\vect{R} + \sum_{i \in \mathcal{I}}  \boldsymbol{\Sigma}_i)^{-1} \right) \quad \textrm{as} \,\,\, \mathcal{P}_t \rightarrow \infty,
\end{split}
\end{equation*}
The first expression above is obtained by applying the Woodbury matrix identity to \eqref{eq:MSE_MMSE}. Equivalently, for the diagonalized estimator we only need to consider $\vect{R}_{\mathrm{diag}}$ and $\vect{S}_{\mathrm{diag}}$ instead of $\vect{R}$ and $\vect{S}$ in the above equations which results in \eqref{eq:diagonalized error floor} as the MSE floor.
\end{IEEEproof}

This proposition shows that the MMSE estimator and the diagonalized estimator exhibit non-zero error floors in the high-power regime. The error floors in \eqref{eq:MMSE error floor} and \eqref{eq:diagonalized error floor} are characterized by the covariance matrix of the own channel and the interfering channels. Clearly, the pilot contamination is the cause of the error floor, which explains the fundamental difference from the noise-limited case where the MSEs approached zero asymptotically.

The next proposition shows that the PEACH and W-PEACH estimators also exhibit MSE floors under pilot contamination.

\begin{proposition}\label{proposition:Pilot contamination PEACHs}
As the pilot power $\mathcal{P}_t \rightarrow \infty$ with the pilot matrix $\vect{P} = \sqrt{\mathcal{P}_t} \vect{I}$, the MSE of PEACH and W-PEACH estimators
converge to the non-zero MSE floors
\begin{equation}\label{eq:Pilot_contamination_PEACH}
\tr \big(\vect{R} +   \vect{R}  \vect{B}_L (\vect{R} + \sum_{i \in \mathcal{I}}  \boldsymbol{\Sigma}_i ) \vect{B}_L  \vect{R} - 2   \vect{R}  \vect{B}_L \vect{R} \big)
\end{equation}
and
\begin{equation}\label{eq:Pilot_contamination_WPEACH}
\tr\left( \vect{R} - \vect{\tilde{b}}^H \vect{\tilde{A}}^{-1} \vect{\tilde{b}}\right)
\end{equation}
respectively, where $\Lambda \! = \! \max_n \lambda_n(\vect{R} \!+\! \sum_{i \in \mathcal{I}}  \boldsymbol{\Sigma}_i) + \min_n \lambda_n(\vect{R} \!+\! \sum_{i \in \mathcal{I}}  \boldsymbol{\Sigma}_i)$, $\vect{B}_L = \frac{2}{\Lambda} \sum_{l=0}^{L} \big( \vect{I} -  \frac{2}{\Lambda}  (\vect{R} + \sum_{i \in \mathcal{I}}  \boldsymbol{\Sigma}_i) \big)^{l}$, $[\vect{\tilde{A}}]_{ij} = \alpha_{\mathrm{w}}^{i+j}  \tr \left( \vect{R}^2 (\vect{R} + \sum_{i \in \mathcal{I}}  \boldsymbol{\Sigma}_i)^{i+j-1} \right)$ and $[\vect{\tilde{b}}]_i = \alpha_{\mathrm{w}}^i  \tr \left( \vect{R}^2  (\vect{R} + \sum_{i \in \mathcal{I}}  \boldsymbol{\Sigma}_i)^{i-1} \right)$.
\end{proposition}
\begin{IEEEproof}
The proof is similar to Proposition \ref{proposition:Noise-limited}. In this case, the MSE expression in \eqref{eq:MSE_PEACH} for PEACH is rewritten as
\begin{equation} \label{eq:MSE-pilotcont-PEACH}
\tr \big(\vect{R} \!+ \!  \vect{R}  (\mathcal{P}_t \vect{A}_L) (\vect{R} \!+\! \sum_{i \in \mathcal{I}}  \boldsymbol{\Sigma}_i \!+\! \frac{\sigma^2}{\mathcal{P}_t} \vect{I}) (\mathcal{P}_t \vect{A}_L^H)  \vect{R} \!-\! 2   \vect{R}   (\mathcal{P}_t \vect{A}_L)  \vect{R} \big)
\end{equation}
where $\mathcal{P}_t \vect{A}_L =  \sum_{l=0}^L  \mathcal{P}_t \alpha  \big( \vect{I} - \mathcal{P}_t \alpha (\vect{R} + \sum_{i \in \mathcal{I}}  \boldsymbol{\Sigma}_i + \frac{\sigma^2}{\mathcal{P}_t} \vect{I}) \big)^l \rightarrow \vect{B}_L$.  This is due to the fact that $\frac{\sigma^2}{\mathcal{P}_t} \vect{I} \rightarrow 0$  and $\mathcal{P}_t \alpha  \rightarrow \frac{2}{\Lambda}$ as $\mathcal{P}_t \rightarrow \infty$, where $\Lambda  \! = \! \max_n \lambda_n(\vect{R} \!+\! \sum_{i \in \mathcal{I}}  \boldsymbol{\Sigma}_i) + \min_n \lambda_n(\vect{R} \!+\! \sum_{i \in \mathcal{I}}  \boldsymbol{\Sigma}_i)$ . By considering all these limits, the MSE in \eqref{eq:MSE-pilotcont-PEACH} converges to the non-zero MSE floor \eqref{eq:Pilot_contamination_PEACH}.

Also, for W-PEACH, we follow the similar approach where the limits of each element of $\vect{DAD}$ and $\vect{Db}$ as $\mathcal{P}_t \rightarrow \infty$ are given by
\begin{equation}
[\vect{DAD}]_{ij} \rightarrow  \alpha_{\mathrm{w}}^{i+j}  \tr \big( \vect{R}^2 (\vect{R} + \sum_{i \in \mathcal{I}}  \boldsymbol{\Sigma}_i)^{i+j-1} \big)
\end{equation}
and
\begin{equation}
[\vect{Db}]_i \rightarrow \alpha_{\mathrm{w}}^i  \tr \big( \vect{R}^2  (\vect{R} + \sum_{i \in \mathcal{I}}  \boldsymbol{\Sigma}_i)^{i-1} \big).
\end{equation}
As in Proposition \ref{proposition:Noise-limited}, it is concluded that $\vect{DAD}$  and $\vect{Db}$ converge to $\vect{\tilde{A}}$ and $\vect{\tilde{b}}$ in the limit which results in $\vect{b}^H \vect{A}^{-1} \vect{b} = \vect{\tilde{b}}^H \vect{\tilde{A}}^{-1} \vect{\tilde{b}}$. Then, it is easily shown that the MSE expression \eqref{eq:minimum MSE} converges to the non-zero floor \eqref{eq:Pilot_contamination_WPEACH}, which is a function of the covariance matrices of the desired and interfering channels, but not the pilot power or noise power.
\end{IEEEproof}

We conclude that the performance of all of the estimators (i.e., the conventional MMSE and the proposed diagonalized, PEACH and W-PEACH estimators) saturate as the pilot power grows large under pilot contamination. This is an expected result for the PEACH estimators, for which the MSEs saturated also in the noise-limited case, while the saturation for the MMSE and diagonalized estimators is completely due to pilot contamination.

\subsection{Numerical Examples}
\label{subsection:examples}

To evaluate the performance of our proposed estimators, we consider a large-scale MIMO system with $N_r = 100$ and $N_t = 10$ antennas and the pilot length $B = 10$. Without loss of generality, we assume zero-mean channel and disturbance, since the non-zero mean assumption has no impact on the MSE performance as shown earlier in Section \ref{section:Low_complexity estimators}. We follow the Kronecker model \cite{Shiu2000a} to describe correlation among antennas of the desired and disturbance MIMO channels. In the simulation, the covariance matrix of a MIMO channel is modeled as $\vect{R} = \vect{R_t} \otimes \vect{R_r}$, where $\vect{R_t} \in \mathbb{C}^{N_t \times N_t}$ and $\vect{R_r} \in \mathbb{C}^{N_r \times N_r}$ are the spatial covariance matrices at the transmitter and receiver sides, respectively. Following the same modeling, we have $\boldsymbol{\Sigma}_i  = \sqrt{\beta_i} \boldsymbol{\Sigma}_{\vect{t}_i}  \otimes \sqrt{\beta_i} \boldsymbol{\Sigma}_{\vect{r}_i}$ for $i \in \mathcal{I}$ where the covariance matrices are weakened by the factor $ \beta_i \geq 0$. This factor represents how severe the pilot contamination part is: $\beta_i=0$ represents the noise-limited case, while $\beta_i=1$ represents the case when the useful channel and the $i$th interfering channel are equally strong.

To generate covariance matrices, we use the exponential correlation model from \cite{Loyka2001a}. All the covariance matrices have diagonal elements equal to one which results in $\tr( \vect{R} ) = N_tN_r$ and $\tr( \boldsymbol{\Sigma}_i ) = \beta_iN_tN_r$. We assume that there are two dominating interfering cells, $i=1,2$. The correlation coefficients for the spatial covariance matrices $\vect{R_t}$, $\vect{R_r}$, $\boldsymbol{\Sigma}_{\vect{t}_i}$ and $\boldsymbol{\Sigma}_{\vect{t}_i}$ where $i=1,2$ are as follows, respectively:
\begin{align*}
r_t &= 0.4 \cdot e^{-j0.9349\pi},  &r_r = 0.9 \cdot e^{-j0.9289\pi}, \\
\sigma_{t,1} &= 0.35 \cdot  e^{-j0.8537\pi},  &\sigma_{r,1} = 0.9 \cdot e^{-j0.7464\pi}, \\
\sigma_{t,2} &= 0.4 \cdot e^{-j0.4583\pi},  &\sigma_{r,2} = 0.9 \cdot e^{-j0.2649\pi}.
\end{align*}
Note that the phases for the correlation coefficients can be chosen randomly, but describe certain channel directivity. We define the normalized pilot SNR as $\gamma = \frac{ \mathcal{P}_t }{ \sigma^2 }$
where $ \mathcal{P}_t = \frac{1}{B} \tr( \vect{P}^H \vect{P} )$ is the average pilot power.

We use the normalized MSE, defined as $\frac{ \textrm{MSE}}{ \tr (\vect{R}) }$, as the performance measure.
In all the figures, we compare the performance of the proposed estimators with the conventional MMSE and MVU estimators. The pilot matrix is $ \vect{P} = \sqrt{\mathcal{P}_t} \vect{I}$. In \cite{Shariati2014a}, it has been shown that this choice of pilot matrix, i.e., the scaled identity, performs (in the MSE sense) almost identical to the optimally robust designed pilot when the channel covariance matrix is uncertain and this uncertainty is bounded by using some norm constraints.

\begin{figure*}[htp]
\centering
\subfigure[$\beta = 0$: Noise-limited scenario.]{\includegraphics[width=0.6\columnwidth]{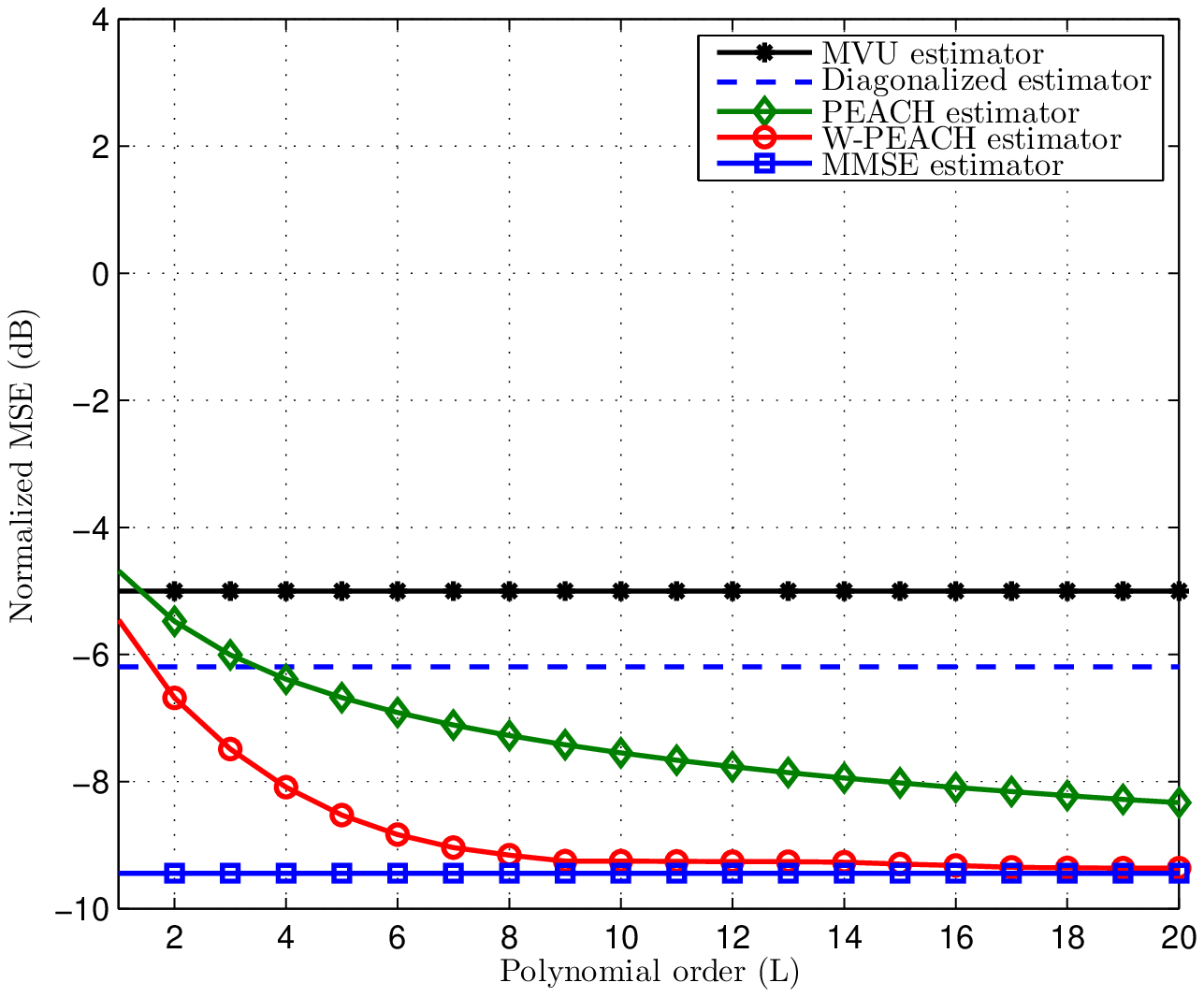}}\quad
\subfigure[$\beta = 0.1$: Pilot contaminated scenario.]{\includegraphics[width=0.6\columnwidth]{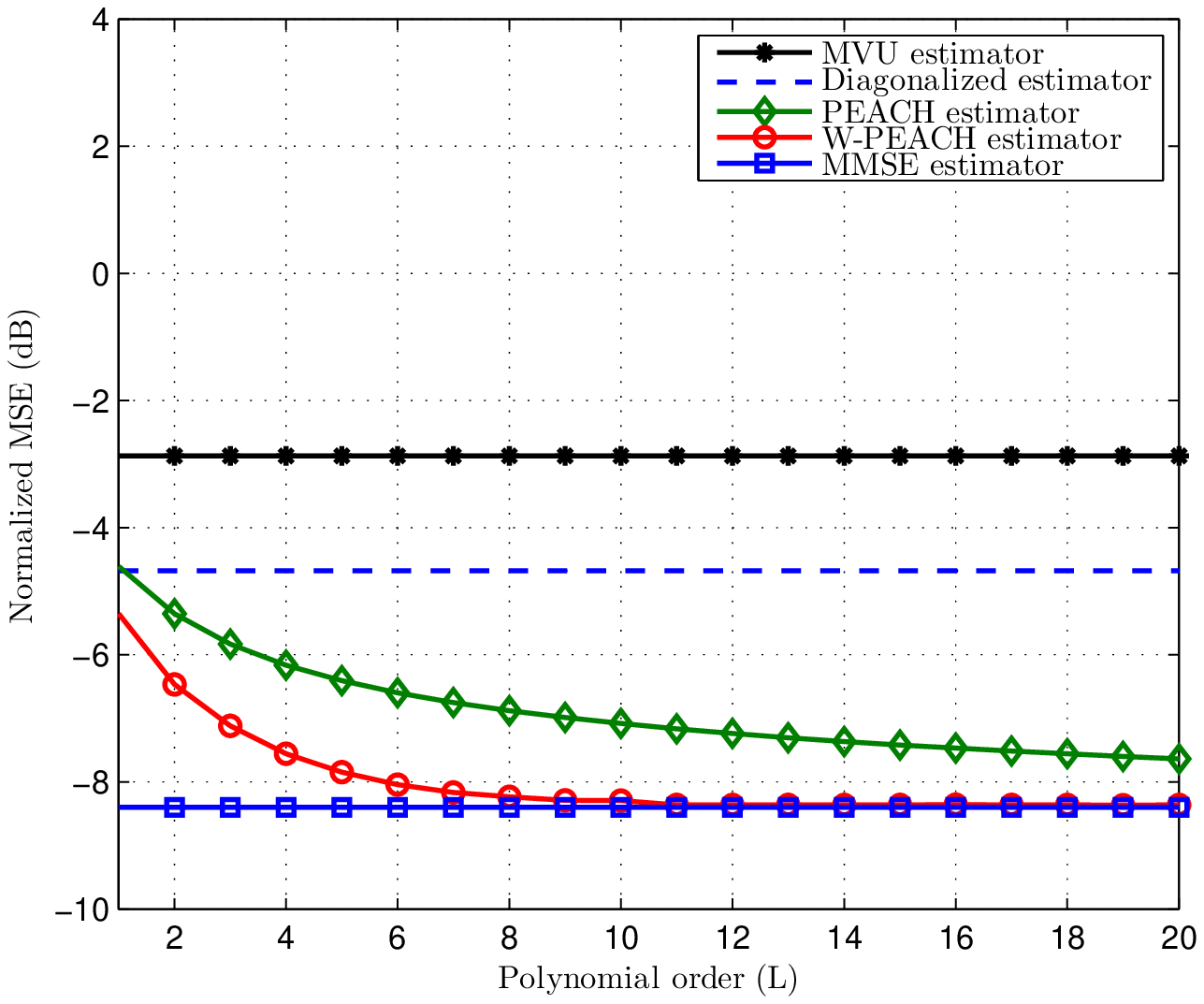}}
\subfigure[$\beta = 1$: Pilot contaminated scenario.]{\includegraphics[width=0.6\columnwidth]{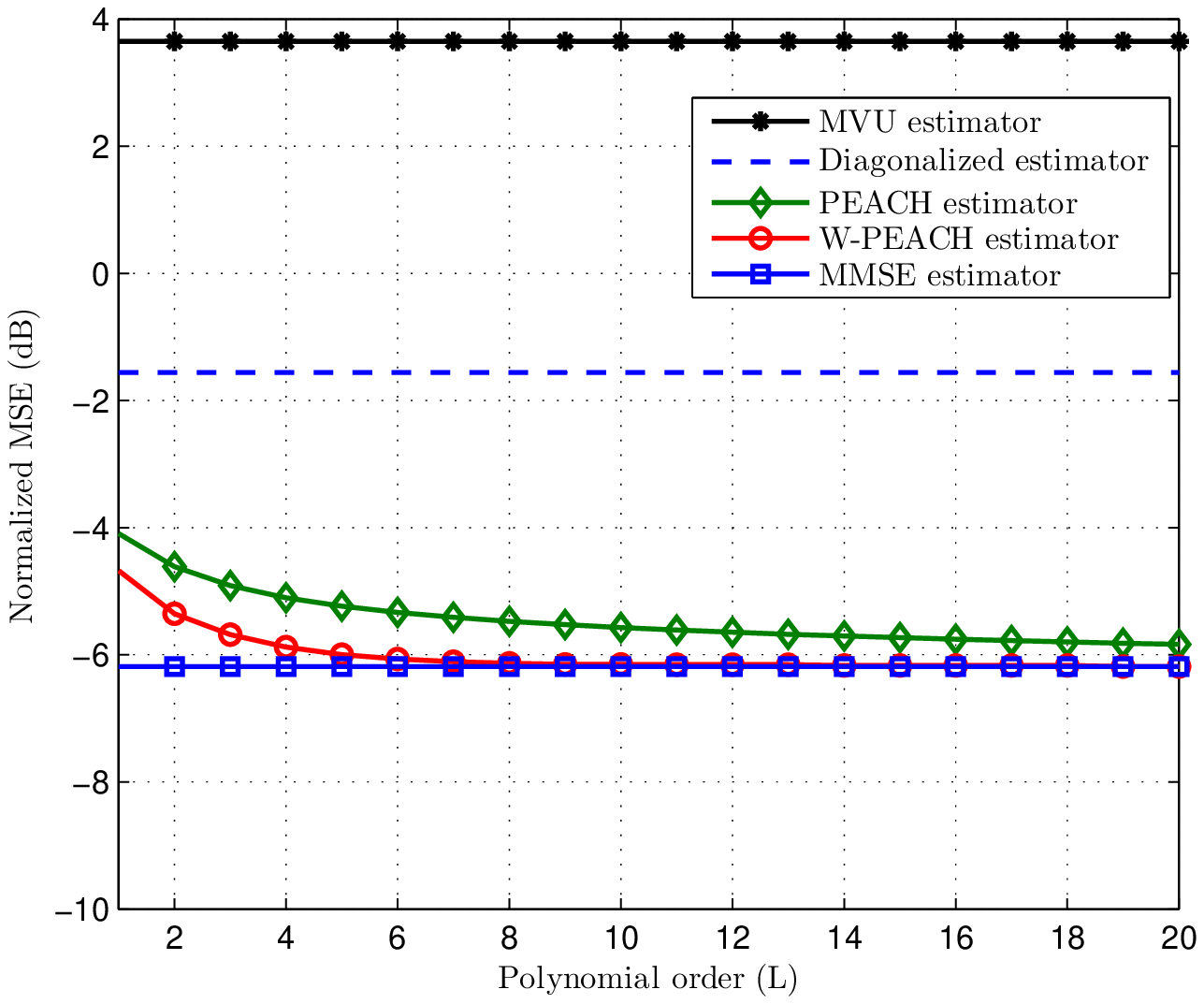}}
\caption{MSE comparison of different estimators as a function of the polynomial degree $L$ for different interference scenarios.}
\label{fig:Fig1}
\end{figure*}

In Fig.~\ref{fig:Fig1}, the MSE has been plotted as a function of the polynomial degree $L$. The noise-limited scenario is given by $\beta = 0$, while $\beta=0.1$ and $\beta=1$ (we assume that $\beta_1=\beta_2=\beta$) represent the scenarios when the two interfering cells have interfering channels which are $10$ dB weaker than or equally strong as the desired channel, respectively. The SNR is $\gamma = 5$ dB. As can be seen from Fig.~\ref{fig:Fig1}, the MSEs of both PEACH and W-PEACH estimators decrease when increasing $L$.
Interestingly, W-PEACH approaches the MSE-values of the MMSE estimator very quickly, while PEACH needs a higher $L$ than W-PEACH to get close to the MMSE curves. The W-PEACH estimator outperforms the MVU, diagonalized and PEACH estimators in all interference scenarios for any value of $L$. Whereas the PEACH estimator outperforms the MVU and the diagonalized estimators under pilot contamination, i.e., $\beta \neq 0$, and outperforms them for $L \geq 2$ and $L \geq 4$, respectively, in the noise-limited case. It is concluded that W-PEACH is near-optimal at quite small $L$, and that PEACH and W-PEACH estimators achieve a better performance than the diagonalized estimator even for small $L$.

\begin{figure*}[htp]
\centering
\subfigure[$\beta = 0$: Noise-limited scenario.]{\includegraphics[width=0.6\columnwidth]{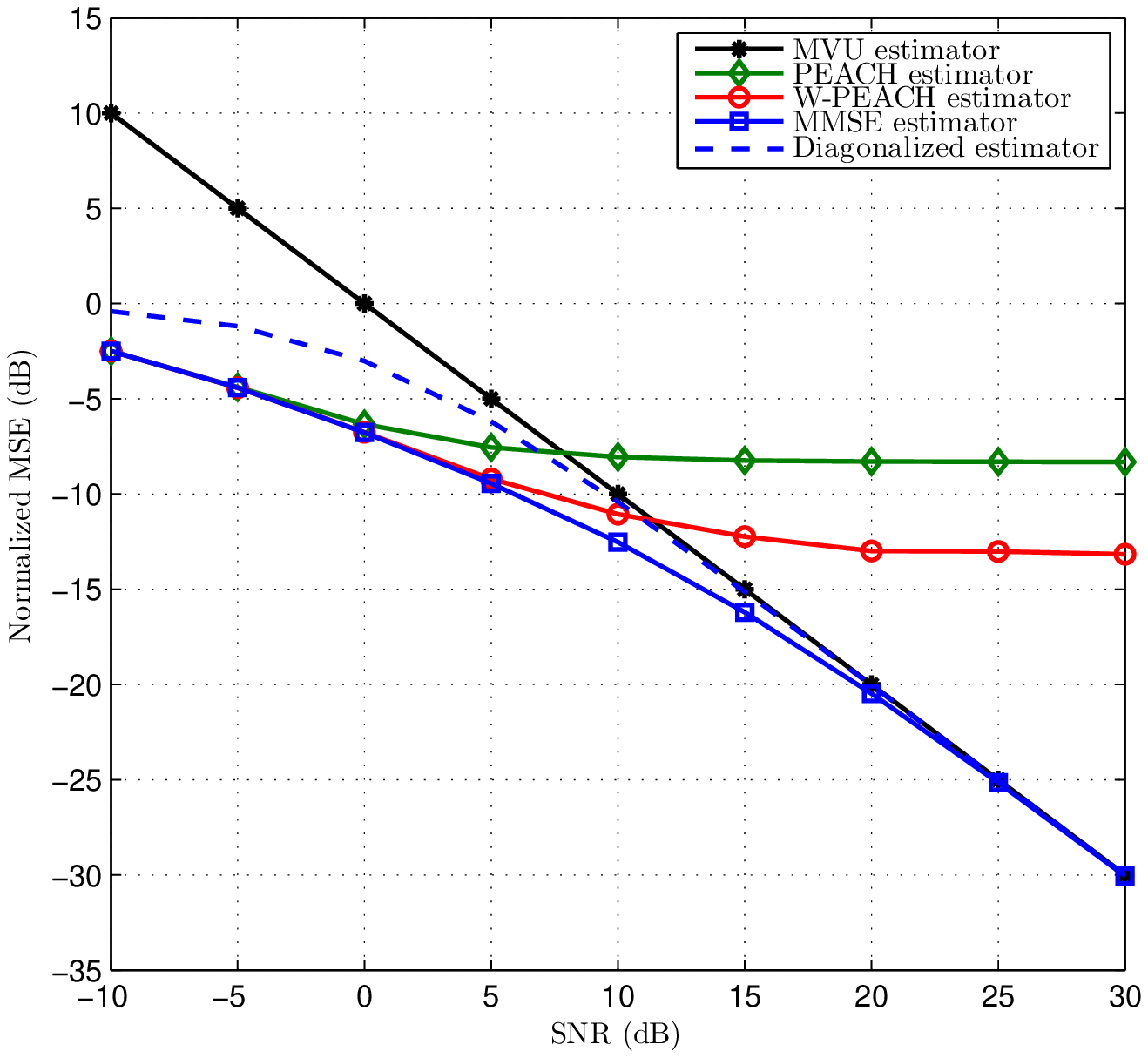}}\quad
\label{subfig:Fig2a}
\subfigure[$\beta = 0.1$: Pilot contaminated scenario.]{\includegraphics[width=0.6\columnwidth]{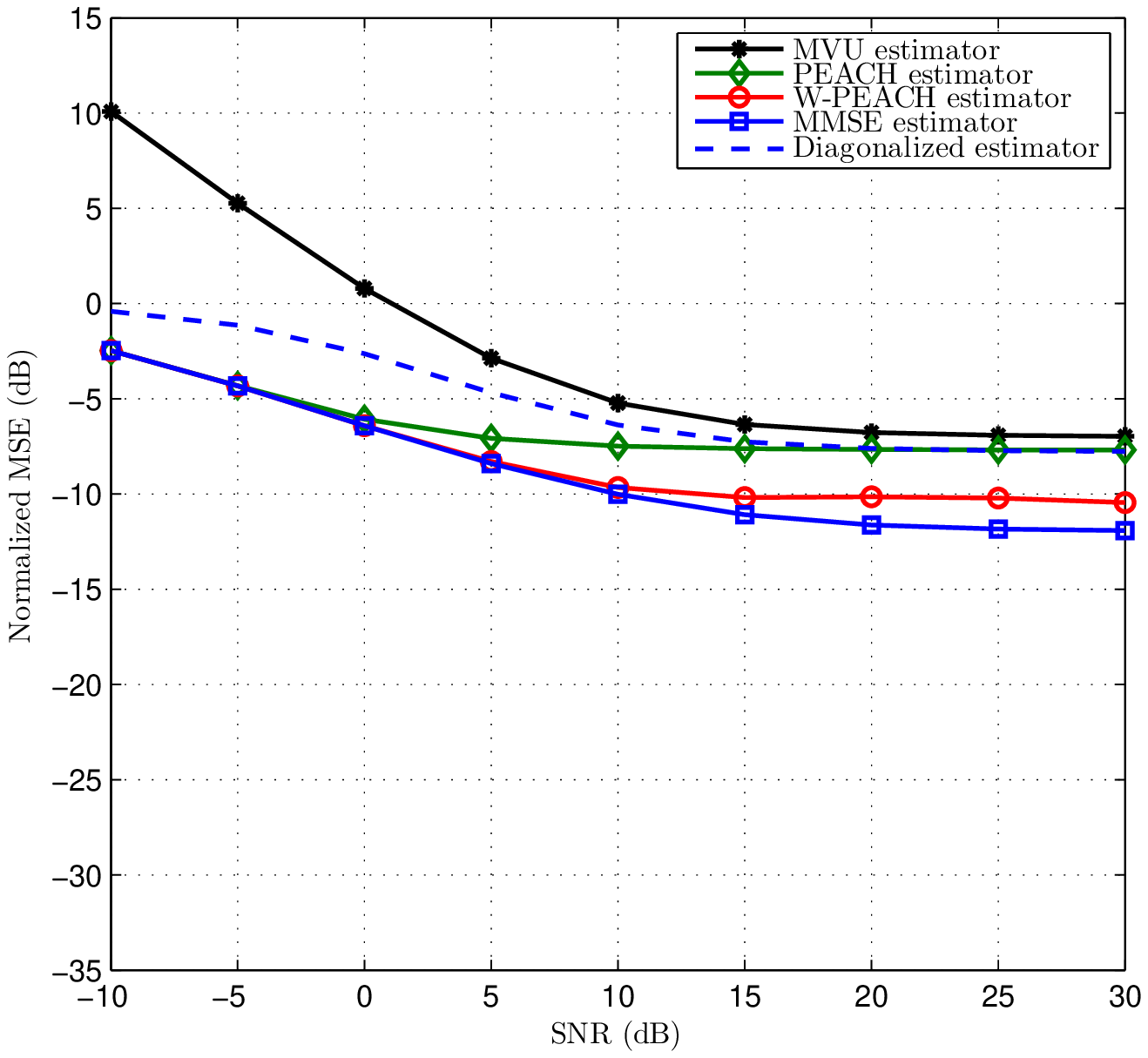}}
\label{subfig:Fig2b}
\subfigure[$\beta = 1$: Pilot contaminated scenario.]{\includegraphics[width=0.6\columnwidth]{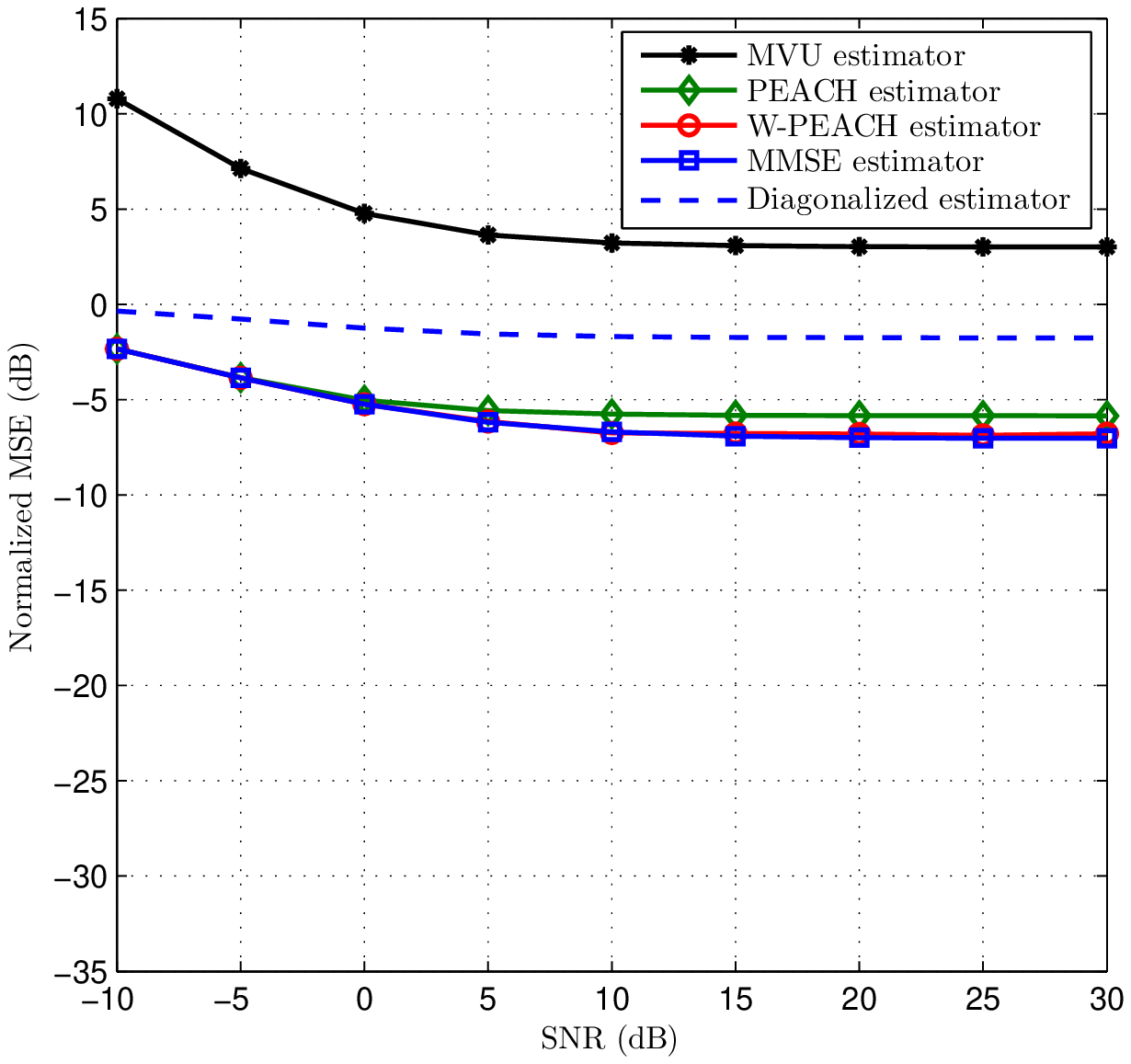}}\quad
\label{subfig:Fig3a}
\caption{MSE comparison of different estimators as a function of SNR $\gamma$ for different interference scenarios.}
\label{fig:Fig2}
\end{figure*}

In Fig.~\ref{fig:Fig2}, we compare different estimators with or without additional interference from pilot contamination. We consider a fixed $L=10$ and vary the SNR $\gamma$. As expected, the MSEs of MMSE, diagonalized and MVU estimators decay steeply to zero when the $\gamma$ increases in the noise-limited scenario. However, as proved in Proposition \ref{proposition:Noise-limited}, the MSEs of PEACH and W-PEACH saturate to non-zero error floors. Under pilot contamination (i.e., $\beta \neq 0$) the performance of all these estimators converge to non-zero error floors. This observation comply with the results stated in Propositions \ref{proposition:Pilot contamination MSE floor} and \ref{proposition:Pilot contamination PEACHs}.  This behavior can be interpreted from another view point. The MSE values are affected by another feature of the system: \textit{signal-to-interference-and-noise ratio (SINR)}.\footnote{The SINR is intimately connected to the MSE. For example, we have $\textrm{MSE} \geq \frac{N_t N_r}{1+\mathrm{SINR}}$ in the special case of $ \vect{P} = \sqrt{\mathcal{P}_t} \vect{I}$, $\vect{R}=\vect{I}$, and $\boldsymbol{\Sigma}_i = \beta_i \vect{I}$. Equality is then achieved by the MMSE estimator.
In general, the SINR needs to grow asymptomatically to infinity if the MSE should approach zero.} Under pilot contamination, the SINR converges to a constant as $\gamma$ increases. More specifically, note that the SINR (when $B = N_t$) is defined as
\begin{equation} \label{eq:SINR-expression}
\mathrm{SINR} = \frac{\mathbb{E}\{ \| \widetilde{\vect{P}} \vect{h} \|^2 \} }{\mathbb{E}\{ \|\vect{n}\|^2 \}} = \frac{ \mathcal{P}_t}{ \sigma^2 + \mathcal{P}_t K \beta}=\frac{\gamma}{1 + \gamma K \beta }
\end{equation}
where $K$ is the number of interferers. As $\gamma$ increases, the SINR in \eqref{eq:SINR-expression} approaches $\frac{1}{K \beta}>0$, thus making the MSEs approach some non-zero limits and become independent of the pilot power $\mathcal{P}_t$.

We observe from Fig.~\ref{fig:Fig2} that pilot contamination only has a small impact on the PEACH and W-PEACH estimators;
in fact, pilot contamination is beneficial in the sense that it reduces the gap to the optimal MMSE estimator; for example, when $\beta=1$ the performance of W-PEACH estimator is identical to that of the MMSE estimator. This important result shows that PEACH estimators are near-optimal in realistic scenarios. The result is explained as follows. For any fixed $L$, PEACH and W-PEACH converge to a non-zero MSE when $\gamma$ increases, due to the bias generated by the approximation error. Since this also happens for the MMSE and MVU estimators under pilot contamination, the relative loss of using the proposed low-complexity estimators is smaller. Consequently, we can reduce $L$ as $\beta$ increases and still achieve near-optimal performance.

In terms of computational complexity, we note that the MVU estimator has the same low complexity as the proposed diagonalized estimator in the noise-limited scenario and for the scaled identity pilot matrix. However, Fig.~\ref{fig:Fig1} and Fig.~\ref{fig:Fig2} show that the diagonalized estimator always outperform the MVU estimator. This is because the diagonalized estimator exploits parts of the channel statistics.

Another interesting observation from Fig.~\ref{fig:Fig2} is how differently the diagonalized estimator performs in different interference scenarios and SNR ranges. The MSE tends to zero in the noise-limited scenario. This implies that there is little loss of using the simple diagonalized estimator at high SNRs since the estimator does not need the spatial correlation to achieve low MSEs in this SNR regime. Hence, the PEACH estimators are only useful at low and medium SNRs in the noise-limited case. However, in the pilot contaminated case the PEACH estimators have a performance advantage throughout the whole SNR range.

\begin{figure}
   \begin{center}
   \includegraphics[width=\columnwidth]{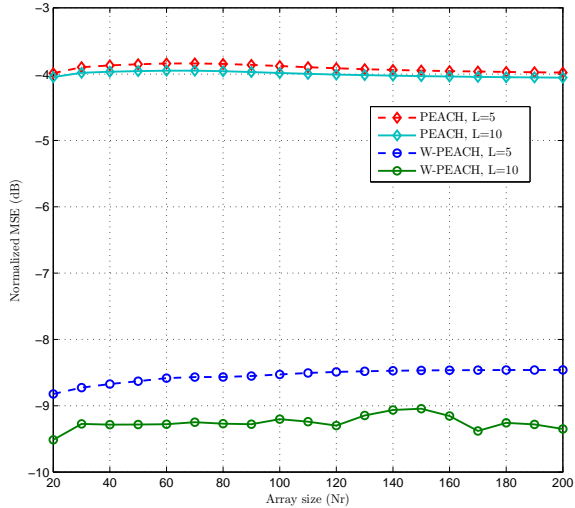}\\ \vskip-3mm
 \caption{Normalized performance of PEACH and W-PEACH estimators for different number of receive antennas.}
   \label{fig:Fig3}
   \end{center}  \vskip-5mm
\end{figure}

In order to illustrate that the estimation performance of the proposed PEACH estimators does not scale with the number of antennas for fixed $L$, we plot in Fig.~\ref{fig:Fig3} the MSE of PEACH and W-PEACH for different number of receive antennas $N_r$ while $N_t$ is fixed to $10$. From Fig.~\ref{fig:Fig3}, we conclude that for a given $L$, there is a certain level of approximation accuracy for the matrix inversion and it determines the MSE performance while there is no clear dependence on the channel dimensions. This result complies with the reasoning in Section \ref{section:Low_complexity estimators} related to Lemma \ref{lemma:inversion-expansion}, as well as the corresponding results in the detection literature \cite{Honig2001a}. This property is indeed one of the main benefits of the PEACH estimators.

\begin{figure}
   \begin{center}
   \includegraphics[width=\columnwidth]{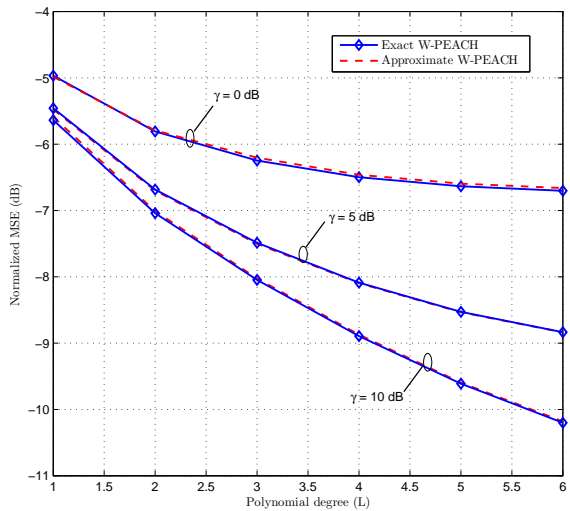}\\ \vskip-3mm
 \caption{Comparison of W-PEACH estimator and Approximate W-PEACH estimator in a noise-limited scenario ($\beta=0$) for different SNR $\gamma$ values.}
   \label{fig:Fig4}
   \end{center}  \vskip-5mm
\end{figure}
\begin{figure}
   \begin{center}
   \includegraphics[width=\columnwidth]{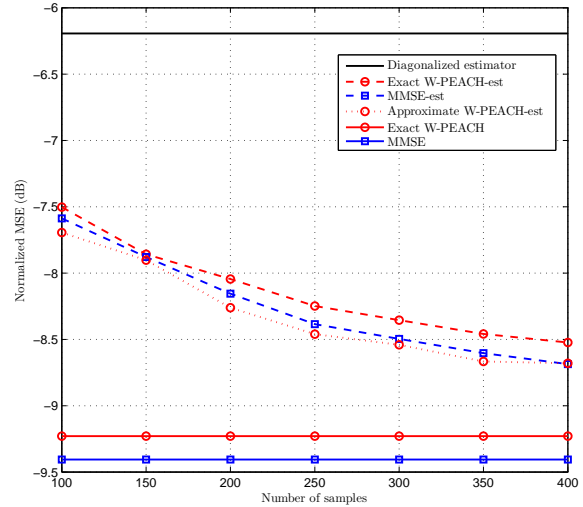}\\ \vskip-3mm
 \caption{Performance comparison of different estimators using the true and sample covariance matrices.}
   \label{fig:Fig5}
   \end{center}  \vskip-5mm
\end{figure}

Next, we focus on the low-complexity approach in Algorithm \ref{algorithm_low-complexity} for finding the weights. First, in Fig.~\ref{fig:Fig4} we illustrate how the approximate weights compared to the optimal weights perform when the perfect covariance matrices are available. Then, in Fig.~\ref{fig:Fig5} we investigate what happens if we only have an imperfect estimate $\hat{\vect{R}}$ of the channel covariance matrix using some finite number of samples $N \leq N_tN_r$. Fig.~\ref{fig:Fig4} considers a noise-limited scenario and a time window of length $T=100$. Although $T \ll B N_r$, we observe that the approximate W-PEACH estimator which exploits the approximate weights from Algorithm \ref{algorithm_low-complexity} gives almost identical performance as the W-PEACH estimator with optimal weights computed according to Theorem \ref{theorem:MSE-minimizing-weights}. This confirms that the W-PEACH estimator is indeed a low-complexity channel estimator suitable for large-scale MIMO systems.


All the simulations so far are done under the assumption that the covariance matrices are perfectly known at the receiver. Next, in Fig.~\ref{fig:Fig5}, we study how imperfect statistical information affects the performance of the MMSE and W-PEACH estimators. For this numerical example, we consider a noise-limited scenario with $N_t = 4$, $N_r = 100$, $L=8$, and $\gamma = 5$ dB. Note that in large-scale noise-limited cases, the noise variance $\sigma^2$ can be easily obtained. However, it is important to evaluate how sensitive the estimators are to imperfect channel statistics. In this figure, we compare the different estimators. The curves marked by $\mathrm{-est}$ at the end of their names are based on the estimated covariance matrix $\hat{\vect{R}}$ described in Section \ref{subsec:sample-covariance}, where the optimal parameter $\kappa^\star$ is obtained using Theorem \ref{theorem:kappa-opt}. The other curves are based on the true covariance matrix $\vect{R}$. Fig.~\ref{fig:Fig5} shows that even for number of samples $N$ smaller than the matrix dimension $N_t N_r$, we can achieve a reasonably \textit{good} performance using $\hat{\vect{R}}$ (recall that it is an affine function of the sample covariance matrix). Moreover, it is shown that the proposed W-PEACH estimator, either using its optimal weights from Theorem \ref{theorem:MSE-minimizing-weights} (Exact W-PEACH) or approximate weights from Algorithm \ref{algorithm_low-complexity} (Approximate W-PEACH), is robust to the statistical uncertainty and performs close to the MMSE estimator.
As expected, it is also observed that using Algorithm \ref{algorithm_low-complexity}, we are able to track the channel's variations better which results in a superior performance as compared to MMSE-est and Exact W-PEACH-est. Observe that the W-PEACH estimator clearly outperforms the diagonalized estimator, implying that we gain from exploiting some of the spatial correlation even when the channel covariance matrix is not perfectly known.

\begin{figure*}[htp]
\centering
\subfigure[Number of received antennas $(N_r)$]{\includegraphics[width=\columnwidth]{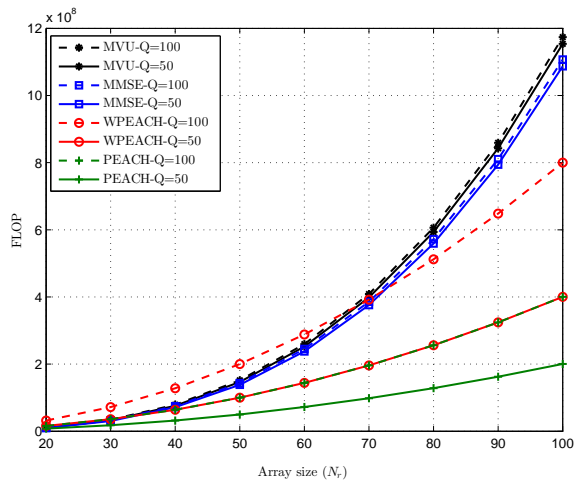}}\quad
\label{subfig:Fig6a}
\subfigure[Number of received antennas $(N_r)$]{\includegraphics[width=\columnwidth]{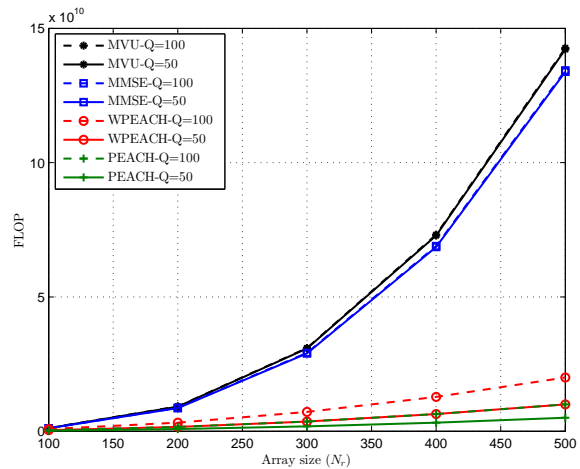}}
\label{subfig:Fig6b}
\caption{Computational Complexity in FLOPs of different channel stationarity conditions $Q$ versus number of received antennas $N_r$ when $L = 2$ and $N_t = B = 10$.}
\label{fig:Fig6}
\end{figure*}

\begin{figure*}[htp]
\centering
\subfigure[Number of received antennas $(N_r)$]{\includegraphics[width=\columnwidth]{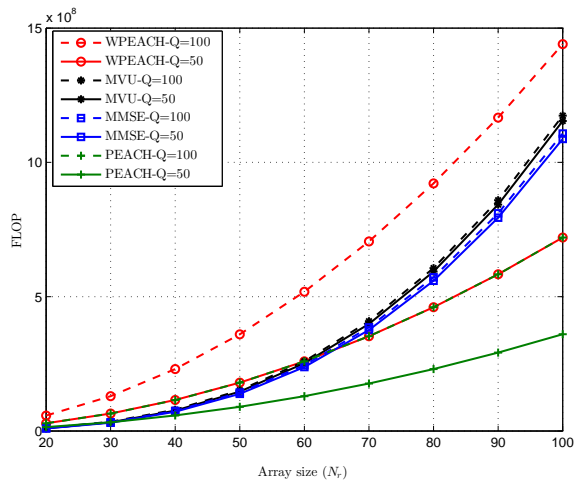}}\quad
\label{subfig:Fig7a}
\subfigure[Number of received antennas $(N_r)$]{\includegraphics[width=\columnwidth]{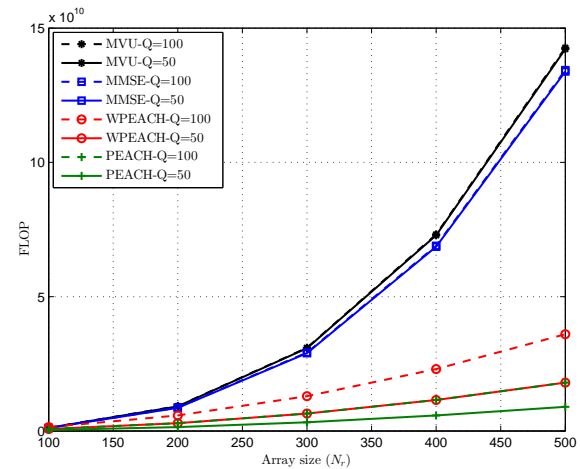}}
\label{subfig:Fig7b}
\caption{Computational Complexity in FLOPs of different channel stationarity conditions $Q$ versus number of received antennas $N_r$ when $L = 4$ and $N_t = B = 10$.}
\label{fig:Fig7}
\end{figure*}

Finally, in the Figs.~\ref{fig:Fig6} and \ref{fig:Fig7} we compare the exact computational complexities of four estimators: MMSE, MVU, PEACH and W-PEACH. In these figures, we plot the number of FLOPs per second versus the number of antennas at the receiver side $N_r$ for different vales of $Q$ (i.e., different stationarity conditions) and different polynomial degrees $L$. We assume $T_{\mathrm{tot}} = \tau_s = 5$ sec. As mentioned in Section \ref{subsec:Summ-Comp-Comp}, these factors affect the exact computational complexity. Observe that the presumed value of $\tau_s$ (or $\tau_c$) change the number of FLOPs but it has no effect on the relative computational complexities of these different estimators. From both figures, we conclude that the PEACH estimator has the lowest computational complexity, which was also proved analytically.

As can be seen in Fig.~\ref{fig:Fig6} for $L = 2$, the W-PEACH estimator has lower complexity than the MMSE estimator when $N_r \geq 35$ for $Q = 50$ and $N_r \geq 73$ for $Q = 100$. However, by increasing the polynomial degree to $L = 4$ (i.e., achieving near-optimal MSEs) a higher number of antennas is needed for W-PEACH estimator: $N_r \geq 135$ to outperform the MMSE estimator in terms of complexity when $Q = 100$, while it is less complex for $N_r \geq 65$ when $Q = 50$. Note that from Fig.~\ref{fig:Fig1} it can be concluded that even with $L = 2$ and $4$, we achieve a reasonably good performance. Also, recall that all the exact complexity analysis is done under the assumption that $\vect{S} \neq \vect{I}$, i.e., pilot contaminated scenario, for which the given values of $L$ provide even better performance compared to the optimal MMSE estimator.

\section{Conclusions}
\label{section:Conclusions}

Large-scale MIMO techniques provide high spatial resolution and array gains, which can be exploited for greatly improved spectral and/or energy efficiency in wireless communication systems. However, achieving  these potential improvements in practice rely on acquiring CSI as precisely as possible. On the other hand, enlarging the array size makes the computational complexity of the signal processing schemes a key challenge. The conventional pilot-based MMSE and MVU channel estimators have a computational complexity unsuitable for such real-time systems. In order to address the complexity issue, we have proposed a set of low-complexity PEACH estimators which are based on approximating the inversion of covariance matrices in the MMSE estimator by an $L$-degree matrix polynomial.

The proposed PEACH estimators converge to the MMSE estimator as $L$ grows large. By deriving the optimal coefficients in the polynomial for any $L$, we can obtain near-optimal MSE performance at small values of $L$. It is shown that $L$ does not scale with the system dimensions, but, in practice, the degree $L$ can be selected to balance between complexity and MSE performance. By performing an exact complexity analysis, we have investigated how the proposed estimator perform compared to the MMSE and MVU estimators from complexity point of view under different assumptions of channel stationarity, the polynomial degree $L$ and number of antennas. The analysis proves that the proposed estimators are beneficial for practically large systems. Numerical results are given for noise-limited scenarios as well as under pilot contamination from pilot reuse in adjacent systems. Although pilot contamination generally creates an MSE floor, it is actually beneficial from a complexity point of view since the proposed estimators achieve good performance at smaller $L$ than in noise-limited scenarios. Furthermore, we introduced the lower-complexity diagonalized estimator. It serves as a viable alternative to PEACH estimators in noise-limited scenarios with high SNRs, whereas PEACH estimators outperform it in the whole SNR range under pilot contamination. By using imperfect channel covariance matrices, we have illustrated numerically that the proposed estimators are robust to statistical uncertainty.

\bibliographystyle{IEEEtran}
\bibliography{IEEEabrv,refs}

\end{document}